\def\version{November 25, 2014}

\documentclass[12pt]{article}
\def\macrosPb{} 
\def\macrosHarxiv{} 
\usepackage[psamsfonts]{amsfonts}
\usepackage{amsmath,amssymb,amsthm}
\usepackage[dvips]{graphicx}
\usepackage{appendix}
\usepackage{bbm} 
\usepackage{amsbsy}
\usepackage{enumerate}
\usepackage{cite}

\ifdefined\macrosPa
  \usepackage[textwidth=465pt,textheight=650pt,centering]{geometry} 
\else\ifdefined\macrosPb
  \usepackage[textwidth=500pt,textheight=650pt,centering]{geometry} 
\fi\fi

\ifdefined\macrosS
  \makeatletter
  
  \def\boldsymbol{\pmb}
  \makeatother

  \usepackage{mathptmx}
  \DeclareMathAlphabet{\mathcal}{OMS}{cmsy}{m}{n}
\fi

\def\UseSection{
        \numberwithin{equation}{section}
	\theoremstyle{plain}
        \newtheorem{theorem}    {Theorem}[section]
        \DefineTheorems 
}

\def\DefineTheorems{
	
	\newtheorem{lemma}      [theorem] {Lemma}
	
	\newtheorem{prop}       [theorem] {Proposition}
	
	\newtheorem{cor}        [theorem] {Corollary}

	\theoremstyle{definition}
	\newtheorem{defn}       [theorem] {Definition}

	\newtheorem{rk} 	[theorem] {Remark}
	\theoremstyle{definition}

}

\newcommand{\bt}   {\begin{theorem}}
\newcommand{\et}   {\end  {theorem}}
\newcommand{\bl}   {\begin{lemma}}
\newcommand{\el}   {\end  {lemma}}
\newcommand{\bp}   {\begin{prop}}
\newcommand{\ep}   {\end  {prop}}
\newcommand{\bc}   {\begin{cor}}
\newcommand{\ec}   {\end  {cor}}
\newcommand{\bd}   {\begin{defn}}
\newcommand{\ed}   {\end  {defn}}

\newcommand{\ba}   {\begin{array}}
\newcommand{\ea}   {\end  {array}}
\newcommand{\be}   {\begin{enumerate}}
\newcommand{\ee}   {\end  {enumerate}}
\newcommand{\bi}   {\begin{itemize}}
\newcommand{\ei}   {\end  {itemize}}

\def\eq#1\en{\begin{equation}#1\end{equation}}  
\def\eqsplit#1\ensplit{
	\begin{equation}\begin{split}#1\end{split}\end{equation}
	}
\def\eqalign#1\enalign{
	\begin{align}#1\end{align}
	}
\def\eqmul#1\enmul{
	\begin{multline}#1\end{multline}
	}
\newcommand{\eqarrstar} {\begin{eqnarray*}} 
\newcommand{\enarrstar} {\end{eqnarray*}} 
\newcommand{\eqarray}   {\begin{eqnarray}} 
\newcommand{\enarray}   {\end{eqnarray}} 
\newcommand{\nnb}	{\nonumber \\} 

\newcommand{\lbeq}[1]  {\label{e:#1}}
\newcommand{\refeq}[1] {\eqref{e:#1}}    

\makeatletter
\newcommand{\labelcounter}[2]{{%
	\stepcounter{#1}
	\protected@write\@auxout{}%
	{\string\newlabel{#2}{{\csname the#1\endcsname}{\thepage}}}%
	{\ref{#2}}
	}}
\makeatother
\newcommand{\Cbold} {{\mathbb C}}  
  
\newcommand{\Ebold} {{\mathbb E}}

\newcommand{\Nbold} {{\mathbb N}}

\newcommand{\Rbold} {{\mathbb R}}

\newcommand{\Zbold} {{\mathbb Z}}

\newcommand{\Ical}   {\mathcal{I}}

\newcommand{\Lcal}   {\mathcal{L}} 
\newcommand{\Mcal}   {\mathcal{M}} 
\newcommand{\Ncal}   {\mathcal{N}}

\newcommand{\Qcal}   {\mathcal{Q}}

\newcommand{\Ucal}   {\mathcal{U}} 
\newcommand{\Vcal}   {\mathcal{V}}

\newcommand{\Zd}    {{ {\Zbold}^d }}

\newcommand{\spose}[1] {{\hbox to 0pt{#1\hss}} }
\newcommand{\ltapprox} {\mathrel{\spose{\lower 3pt\hbox{$\mathchar"218$}}
 \raise 2.0pt\hbox{$\mathchar"13C$}}}
\newcommand{\gtapprox} {\mathrel{\spose{\lower 3pt\hbox{$\mathchar"218$}}
 \raise 2.0pt\hbox{$\mathchar"13E$}}}

\UseSection   
\usepackage[usenames]{color}

\definecolor{at}{rgb}{0.0, 0.5, 0.0} 

\newcommand{\LTsym}{{\rm loc}}
\newcommand{\LT}{{\rm Loc}  }
\newcommand{\Rel}{{\rm Proj}}
\newcommand{\Irr}{(1-{\rm Proj})}

\renewcommand{\to} {\rightarrow}

\newcommand{\R}{\Rbold}
\newcommand{\Z}{\Zbold}

\newcommand{\N}{\Nbold}
\newcommand{\C}{\mathbb{C}}

\newcommand{\Lambdabold}{\boldsymbol{\Lambda}}

\newcommand{\1}{\mathbbm{1}}
\newcommand{\Cbf}{\boldsymbol{C}}

\newcommand{\wbf}{\boldsymbol{w}}

\newcommand{\la}{\langle}
\newcommand{\ra}{\rangle}
\newcommand{\supp}{\mathrm{supp}}

\newcommand{\psib}{\bar\psi}
\newcommand{\ci}{\underline{i}}

\newcommand{\jm}{j_\Omega}

\newcommand{\Ex}{\mathbb{E}}

\newcommand{\Qcalnabla}{\Qcal}

\newcommand{\units}{\Ucal}

\newcommand{\pt}{{\rm pt}}

\newcommand{\Vpt}{V_{\rm pt}}
\newcommand{\gpt}{g_{\mathrm{pt}}}
\newcommand{\nupt}{\nu_{\mathrm{pt}}}
\newcommand{\mupt}{\mu_{\mathrm{pt}}}
\newcommand{\zpt}{z_{\mathrm{pt}}}
\newcommand{\ypt}{y_{\mathrm{pt}}}
\newcommand{\lambdapt}{\lambda_{\mathrm{pt}}}
\newcommand{\lambdaapt}{\lambda^{\pp}_{\mathrm{pt}}}
\newcommand{\lambdabpt}{\lambda^{\qq}_{\mathrm{pt}}}
\newcommand{\lambdaa}{\lambda^{\pp}}
\newcommand{\lambdab}{\lambda^{\qq}}
\newcommand{\qapt}{q^{\pp}_{\mathrm{pt}}}
\newcommand{\qbpt}{q^{\qq}_{\mathrm{pt}}}
\newcommand{\qa}{q^{\pp}}
\newcommand{\qb}{q^{\qq}}
\newcommand{\qpt}{q_{\mathrm{pt}}}

\newcommand{\Vch}{\check{V}}

\newcommand{\gch}{\check{g}}
\newcommand{\zch}{\check{z}}
\newcommand{\much}{\check{\mu}}

\newcommand{\Lcallr}{\stackrel{\leftrightarrow}{\Lcal}}

\newcommand{\gbar}{\bar{g}}

\newcommand{\zbar}{\bar{z}}
\newcommand{\mubar}{\bar{\mu}}

\newcommand{\pp}{a}
\newcommand{\qq}{b}
\newcommand{\sigmaa}{\sigma}
\newcommand{\sigmab}{\bar{\sigma}}

\newcommand{\half}{\textstyle{\frac 12}}

\newcommand{\ddp}[2]{\frac{\partial #1}{\partial #2}}

\newcommand{\phib}{\bar\phi}

\ifdefined\macrosH
  \usepackage{xr-hyper}
  \usepackage{hyperref}
  \hypersetup{hypertexnames=false}
  \hypersetup{colorlinks,citecolor=blue,linkcolor=blue}  

\else\ifdefined\macrosHarxiv
  \usepackage{xr-hyper}
  \usepackage{hyperref}
  \hypersetup{hypertexnames=false}

\else
  \newcommand{\texorpdfstring}[2]{#1}
  \usepackage{xr}
\fi\fi

\title{
  A renormalisation group method.
  \\
  III. Perturbative analysis
}

\author{
  Roland Bauerschmidt\thanks{Department of Mathematics,
    University of British Columbia,
    Vancouver, BC, Canada V6T 1Z2.
    Present address: School of Mathematics,
    Institute for Advanced Study,
    Princeton, NJ 08540 USA.
    E-mail: {\tt brt@math.ias.edu}.},\;
  David C.\ Brydges\thanks{Department of Mathematics,
    University of British Columbia,
    Vancouver, BC, Canada V6T 1Z2.
    E-mail: {\tt db5d@math.ubc.ca}, {\tt slade@math.ubc.ca}.}\;
  and Gordon Slade$^\dagger$}

\date{\version}

\begin{document}

\maketitle

\renewcommand{\Cbf}{C}
\renewcommand{\wbf}{w}

\begin{abstract}
This paper is the third in a series devoted to the development
of a rigorous renormalisation group method for lattice field theories
involving boson fields, fermion fields, or both.
In this paper, we motivate and present a general approach towards
second-order perturbative renormalisation, and apply it to a specific
supersymmetric field theory which represents the continuous-time
weakly self-avoiding walk on $\Zd$.  Our focus is on the critical
dimension $d=4$.  The results include the derivation of the
perturbative flow of the coupling constants, with accompanying
estimates on the coefficients in the flow.  These are essential
results for subsequent application to the 4-dimensional weakly
self-avoiding walk, including a proof of existence of logarithmic
corrections to their critical scaling.  With minor modifications, our
results also apply to the 4-dimensional $n$-component $|\varphi|^4$
spin model.
\end{abstract}

\section{Introduction}
\label{sec:intro}

Within theoretical physics, in the study of critical phenomena, or
quantum field theory, or many-body theory, the calculation of
physically relevant quantities such as critical exponents or particle
mass is routinely carried out in a perturbative fashion.  The
perturbative calculations involve tracking the flow of coupling
constants which parametrise a dynamical system evolving under
renormalisation group transformations.  In this paper, we present a
general formalism for second-order perturbative renormalisation, and
apply it to the continuous-time weakly self-avoiding walk.

This paper is the third in a series devoted to the development of a
rigorous renormalisation group method.  In part~I of the series, we
presented elements of the theory of Gaussian integration and
defined norms and developed
an analysis for performing analysis with Gaussian
integrals involving both boson and fermion fields
\cite{BS-rg-norm}.  In part~II, we defined and analysed a
localisation operator whose purpose is to extract relevant and
marginal directions in the dynamical system defined by the
renormalisation group \cite{BS-rg-loc}.  We now apply the formalism
of parts~I and II to the perturbative analysis of
a specific supersymmetric field theory that
arises as a representation of the continuous-time weakly self-avoiding
walk \cite{BIS09}.
Our development of
perturbation theory makes contact with the standard technology of
Feynman diagrams as it is developed in textbooks on quantum field theory, but
our differences in emphasis prepare the ground for the control of
non-perturbative aspects in parts~IV and V \cite{BS-rg-IE,BS-rg-step}.

The results of this paper are applied in \cite{BBS-saw4-log,BBS-saw4},
in conjunction with \cite{BBS-rg-flow,BS-rg-IE,BS-rg-step}, to the
analysis of the critical two-point function and susceptibility of the
continuous-time weakly self-avoiding walk.  They are also applied in
\cite{BBS-phi4-log} to the analysis of the critical behaviour of the
$4$-dimensional $n$-component $|\varphi|^4$ spin model.  Our emphasis
here is on the \emph{critical dimension} $d=4$, which is more
difficult than dimensions $d >4$.  In this paper, we derive the
second-order perturbative flow of the coupling constants, and prove
accompanying estimates on the coefficients of the flow.  The flow
equations themselves are analysed in \cite{BBS-rg-flow,BBS-saw4}.
While the results of this paper are for the specific supersymmetric
field theory representing the continuous-time self-avoiding walk, the
principles are of wider validity, and apply in particular to the
$n$-component $|\varphi|^4$ model.

The paper is organised as follows.  We begin in Section~\ref{sec:WPjobs}
by motivating and developing a general approach to perturbation theory.
Precise definitions are made in Section~\ref{sec:main-results-focc},
and the main results are stated in Section~\ref{sec:mr}.
Proofs are deferred to Sections~\ref{sec:Vpt}--\ref{sec:Greekpfs}.
In addition, Section~\ref{sec:Greekpfs} contains a definition and
analysis of the
specific finite-range covariance decomposition that is important in and
used throughout
\cite{BBS-saw4-log,BBS-saw4,BS-rg-IE,BS-rg-step}.

\section{Perturbative renormalisation}
\label{sec:WPjobs}

In this section, we present an approach to perturbative renormalisation
that motivates the definitions of Section~\ref{sec:main-results-focc}.
The analysis is \emph{perturbative}, meaning that it is valid as
a formal power series but in this form cannot be controlled uniformly
in the volume.
We do not directly apply the contents of this section elsewhere,
but they help explain why the definitions and results that follow
in Sections~\ref{sec:main-results-focc}--\ref{sec:mr}
are appropriate and useful.
Also, the approach discussed here provides a perspective which guides
related developments in part~IV \cite{BS-rg-IE},
and which together with part~V \cite{BS-rg-step}
lead to remainder estimates that do apply uniformly in the volume.
In particular, the proof of
\cite[Proposition~\ref{IE-prop:h}]{BS-rg-IE}, which goes beyond
formal power series, relies on the
principles presented here.

Given integers $L,N>1$, let $\Lambda = \Zd/L^N\Z$ denote the discrete torus
of period $L^N$.
Recall the definitions of the boson and fermion fields on $\Lambda$,
and of the
combined bo\-son\-ic-fermionic Gaussian integration $\Ex_{\wbf}$ with
covariance $\wbf$, from \cite[Section~\ref{norm-sec:gint}]{BS-rg-norm}
(for notational simplicity we write the bold-face covariances of \cite{BS-rg-norm}
without bold face here).
Recall also the definition of local monomial in
\cite[\eqref{loc-e:Mm}]{BS-rg-loc}.  Suppose we have a vector space
$\Vcal$ of local polynomials in the boson and/or fermion fields, whose
elements are given by linear combinations of local
monomials.
The evaluation of the fields in an element $V$ of
$\Vcal$ at a point $x\in\Lambda$ is
denoted by $V_{x}$,
and $V(X)$ denotes the sum
\begin{equation}
\lbeq{VXdef}
  V(X) = \sum_{x \in X} V_x.
\end{equation}
Supersymmetry plays no role in these considerations, so we do not
assume in this section that the field theory is supersymmetric.  For
simplicity, we assume here that the elements of $\Vcal$ are
translation invariant on $\Lambda$.  Observable terms, which break
translation invariance, are
handled by adapting what we do here to include the projections
$\pi_\varnothing$ and $\pi_*$ as in \eqref{e:Fpi} below.

The main problem we wish to address is the computation of a Gaussian
integral $\Ex_{\wbf} e^{-V_{0}(\Lambda)}$, where $V_{0} \in \Vcal$,
and where $\wbf=\wbf_N$ is a positive-definite covariance matrix
indexed by $\Lambda$ which approximates the inverse lattice Laplacian
$[-\Delta_{\Zd}]^{-1}$ in the infinite volume limit $N \to \infty$.
We will see that divergences arise due to the slow decay of the
covariance, but that perturbative renormalisation leads to expressions
without divergences, provided the coupling constants are allowed to
depend on scale.  We consider the problem now at the level of formal
power series in the coupling constants, working accurately to second
order and with errors of order $O(V_0^3)$.  The notation $O(V_0^{n})$
signifies a series in the coupling constants whose lowest order terms
have degree at least $n$, and we write $\approx$ to denote equality as
formal power series up to an error $O(V_0^3)$.  By expanding the
exponentials, it is easy to verify that
\begin{align}
\label{e:EIappw}
    \Ex_{\wbf}  e^{-V_0}
    & \approx
    e^{ - \Ex_{\wbf} V_0 + \frac 12 \Ex_{\wbf} ( V_0;  V_0) }
    ,
\end{align}
where the second term in the exponent on the right-hand side is
the \emph{truncated expectation} (or \emph{variance})
\begin{equation}
\label{e:Etrunc}
    \Ex_{\wbf} ( V_0;  V_0)
    =
    \Ex_{\wbf}   V_0^2
    -
    (\Ex_{\wbf}  V_0)^2.
\end{equation}
In \eqref{e:EIappw}--\eqref{e:Etrunc},
the abbreviation $V_0=V_0(\Lambda) = \sum_{x \in \Lambda}V_x$
has left the $\Lambda$ dependence implicit.
Equation~\refeq{EIappw} gives the first two terms of the
\emph{cumulant expansion} and \refeq{Etrunc} is also referred to
as an \emph{Ursell function}.

The formula \eqref{e:EIappw} provides a way to perform the integral,
but it is not useful because in the infinite volume limit the
covariance we are interested in decays in dimension $d=4$ as
$|x-y|^{-2}$, which is not summable in $y$, and this leads to
divergent coefficients in \eqref{e:Etrunc}.  For example, suppose that
there just one field, a real boson field $\phi$, and that $V_0 =
V_0(\Lambda) = \sum_{x \in \Lambda} \phi_x^2$.  Evaluation of
\eqref{e:Etrunc} in this case gives $\Ex_{\wbf} ( V_0; V_0) =
|\Lambda|\sum_{x \in \Lambda} \wbf(0,x)^2$.  The volume factor
$|\Lambda|$ is to be expected, but the sum over $x$ is the
\emph{bubble diagram} and diverges in the infinite volume limit when
$d=4$. This is a
symptom of worse divergences that occur at higher order.

A solution to this famous difficulty of infinities plaguing the
functional integrals of physics is provided by the renormalisation
group method.  For the formulation we are using, we decompose the
covariance as a sum ${\wbf} = {\wbf}_N = \sum_{j=1}^N \Cbf_j$.  Then,
as proved in \cite[Proposition~\ref{norm-prop:conv}]{BS-rg-norm}, the
expectation can be performed progressively via iterated convolution:
\begin{equation}
    \label{e:iterated-expectation}
    \Ex_{{\wbf}} e^{-V_{0}(\Lambda)}
    = \Ex_{\Cbf_N} \circ \Ex_{\Cbf_{N-1}}\theta \circ \dots
    \circ \Ex_{\Cbf_1} \theta e^{-V_{0}(\Lambda)},
\end{equation}
with the operator $\theta$ as defined in
\cite[Definition~\ref{norm-def:theta-new}]{BS-rg-norm} and discussed
around \eqref{e:EWick} below.  This is an extension of the elementary
fact that if $X \sim N(0,\sigma_1^2+\sigma_2^2)$ then we can evaluate
$\Ex(f(X))$ progressively as
\begin{equation}
\label{e:EX1X2}
    \Ex(f(X)) = \Ex( \Ex( f(X_1+X_2) \, | \, X_2) ),
\end{equation}
with independent normal random variables $X_1 \sim N(0, \sigma_1^2)$ and
$X_2 \sim N(0,\sigma_2^2)$.

An essential step is to understand the effect of a single expectation
in the iterated expectation \eqref{e:iterated-expectation}.  For this,
we we seek a replacement
\begin{equation}
\label{e:IW}
    I_j(V,\Lambda) = e^{-V(\Lambda)}(1+W_j(V,\Lambda))
\end{equation}
for $e^{-V(\Lambda)}$, with $W_j(V,\Lambda)=\sum_{x\in
\Lambda}W_j(V,x)$ chosen to ensure that the form of $I_j (V,\Lambda)$
remains stable under expectation.
By stability, we mean that given $V_j$, there exists $V_{j+1}$ such that
\begin{equation}
\label{e:Istab}
    \Ex_{\Cbf_{j+1}}\theta I_{j}(V_j,\Lambda) \approx I_{j+1}(V_{j+1},\Lambda)
\end{equation}
is correct to second order when both sides are expressed as
power series in the coupling constants of $V_{j}$. In particular the
coupling constants of $V_{j+1}$ are power series in the coupling
constants of $V_{j}$. The recursive composition of these power series
expresses $V_{j}$ as a series in the coupling constants of $V_{0}$
but, as explained above, this series has bad properties as $j
\rightarrow \infty$.
However, if $V_{j+1}$ is instead expressed as a function of $V_j$
as in \eqref{e:Istab}, then this opens the door to the possibility to
arrange that \refeq{Istab} holds uniformly in $j$ and $N$.
This is one of the great discoveries of theoretical physics---not in
the sense of mathematical proof, but as a highly effective
calculational methodology.  Its first clear exposition in terms of
progressive integration is due to Wilson \cite{Wils83}, following
earlier origins in quantum field theory \cite{GML54}.

We make several definitions, whose utility will become apparent below.
According to
\cite[Proposition~\ref{norm-prop:conv}]{BS-rg-norm}, for a polynomials $A$
in the fields, the Gaussian expectation with covariance $\Cbf$ can
be evaluated using the Laplacian operator
$\frac{1}{2} \Delta_{\Cbf}$, via
$\Ex \theta A = e^{\frac{1}{2} \Delta_{\Cbf}} A$.
For polynomials $A,B$ in the fields, the truncated
expectation is then given by
\begin{equation}
\label{e:EWick-bis}
    \Ex_{\Cbf}(\theta A;\theta B)
    = e^{\frac{1}{2} \Delta_{\Cbf}}
    (AB) -
    \big(e^{\frac{1}{2} \Delta_{\Cbf}}A\big)
    \big(e^{\frac{1}{2} \Delta_{\Cbf}}B \big)
.
\end{equation}
Given $A,B$, we define
\begin{align}
\label{e:FCAB-H}
    F_{\Cbf}(A,B) &
    = e^{\frac{1}{2} \Delta_{\Cbf}}
    \big(e^{-\frac{1}{2} \Delta_{\Cbf}}A\big)
    \big(e^{-\frac{1}{2} \Delta_{\Cbf}}B \big) - AB,
\end{align}
and conclude that
\begin{equation}
\label{e:EFABC}
    \Ex_{\Cbf}(\theta A;\theta B)
    = F_{\Cbf}(\Ex_{\Cbf}\theta A,\Ex_{\Cbf}\theta  B).
\end{equation}
Also, for $X \subset \Lambda$,
we define $W_j(V,X)= \sum_{x \in X} W_j(V,x)$ with
\begin{equation}
\label{e:WLTF-bis}
    W_j(V,x) = \frac 12 (1-\LT_{x}) F_{ \wbf_j}(V_x,V(\Lambda)),
\end{equation}
with $\LT$ the operator studied in \cite{BS-rg-loc},
and we use this to define $W_j(V,\Lambda)$ in \eqref{e:IW}.
Then we define $P(X)=P(V,X)=\sum_{x\in X}P_x$ by
\begin{equation}
\label{e:PdefF-bis}
    P_x = \LT_x \Ex_{\Cbf_{j+1}}\theta W_j(V,x)
    + \frac 12 \LT_x
    F_{ \Cbf_{j+1}}
    (\Ex_{\Cbf_{j+1}}\theta V_x,\Ex_{\Cbf_{j+1}}\theta V(\Lambda)).
\end{equation}
Finally, the local polynomial $\Vpt$ is defined in terms of $V$ by
\begin{equation}
    \label{e:beta-function2-H}
    \Vpt = \Ex_{\Cbf_{j+1}}\theta V - P.
\end{equation}
In Proposition~\ref{prop:Vptg} below, we present $\Vpt$ in full detail
for the weakly self-avoiding walk.

The following is a version of \cite[Proposition~7.1]{BS11},
with the observables omitted.
Proposition~\ref{prop:I-action} shows that the definitions above lead to a form of the
interaction which is stable in the sense of \eqref{e:Istab}.
Its proof provides motivation for the definitions of $W$ and $\Vpt$
made above.

\begin{prop}\label{prop:I-action}
As formal power series in $V$,
\begin{equation}
    \label{e:I-invariance-H}
    \Ebold_{\Cbf_{j+1}}\theta I_{j} (V,\Lambda)
    \approx
    I_{j+1} (\Vpt,\Lambda)
    ,
\end{equation}
with an error which is $O(V^3)$.
\end{prop}

\begin{proof}
The proof includes some motivational remarks that are not
strictly necessary for the proof.
Suppose that $W_j$ is given; the initial condition is $W_0=0$.
We initially treat $W_j$ as an unknown sequence of \emph{quadratic} functionals of
$V$, of order $O(V^2)$,
and we will discover that \eqref{e:WLTF-bis} is a good choice to
achieve \eqref{e:I-invariance-H}.
We write $V=V(\Lambda)$
to simplify the notation.
We use
\begin{equation}
    e^{-V}(1+W)
\approx
    e^{-V+W},
\end{equation}
together with \eqref{e:IW} and \eqref{e:EIappw}, to obtain
\begin{equation}
\label{e:EIapp}
    \Ex_{\Cbf_{j+1}}\theta I_j(V,\Lambda)
    \approx
    e^{ - \Ex_{\Cbf_{j+1}}\theta V}
    \left[ 1  +
    \Ex_{\Cbf_{j+1}} \theta W_j(V,\Lambda)
    + \frac 12 \Ex_{\Cbf_{j+1}} (\theta V; \theta V)
    \right].
\end{equation}
The second-order term $\Ex_{\Cbf_{j+1}} \theta W_j+\frac 12 \Ex_{\Cbf_{j+1}} (\theta V;
\theta V)$ contains contributions which are marginal and relevant for
the dynamical system on the space of functionals of the fields,
generated by the maps $\Ex_{\Cbf_{j+1}}\theta$.

The idea of the renormalisation group is to track the flow explicitly on a
finite-dimensional subspace of the full
space of functionals of the fields. In our case, this subspace
is the space $\Vcal(\Lambda)$
of local polynomials, and we need to project
onto this subspace of marginal and relevant directions.
Call this projection $\Rel$.
Below, we will relate $\Rel$ the
operator $\LT$ of \cite{BS-rg-loc}.  For now, the one assumption
about $\Rel$ we need is that
\begin{equation}
\label{e:IrrRel}
    \Irr \circ \Ex \theta \circ \Rel = 0.
\end{equation}
In other words, integration of relevant or
marginal terms does not produce irrelevant terms, or, to put it
differently,
the space onto which $\Rel$ projects is $\Ex$-invariant.
Then we define
\begin{equation}
\label{e:PW}
    P(\Lambda)
    =\Rel \left( \Ex_{\Cbf_{j+1}}\theta W_j (V,\Lambda) +
    \frac 12 \Ex_{\Cbf_{j+1}}\big(\theta V;\theta V \big) \right) .
\end{equation}
It follows from \eqref{e:EFABC} that
\begin{equation}
    \Ex_{\Cbf}(\theta V(\Lambda); \theta V(\Lambda))
    =
    F_{\Cbf}(\Ex_{\Cbf}\theta V(\Lambda),\Ex_{\Cbf}\theta V(\Lambda)),
\end{equation}
and hence \eqref{e:PW} is consistent with \eqref{e:PdefF-bis} when $\Rel$
is taken to be $\LT$.
We then define
$\Vpt = \Ex_{\Cbf_{j+1}}\theta V - P$ as in \eqref{e:beta-function2-H}.
From \eqref{e:EIapp},
dropping $\Lambda$ from the notation, we now obtain
\begin{align}
    \Ex_{\Cbf_{j+1}} \theta I_j(V)
    & \approx
    e^{-\Vpt}
    \left( 1 + \Irr \left(\Ex_{\Cbf_{j+1}} \theta W_j(V)
    + \frac 12 \Ex_{\Cbf_{j+1}} (\theta V; \theta V)
    \right)\right).
\end{align}
In this way, the effect of the marginal and relevant
terms in \eqref{e:EIapp} has been incorporated into $\Vpt$.

The demand that the form of the interaction remain stable under
expectation now becomes
\begin{equation}
\label{e:Istability}
    \Ex_{\Cbf_{j+1}} \theta I_j(V) \approx e^{-\Vpt}(1+W_{j+1}(\Vpt)),
\end{equation}
with
\begin{equation}
\label{e:Wfind1}
    W_{j+1}(\Vpt) \approx
    \Irr \left(\Ex_{\Cbf_{j+1}} \theta W_j(V)
    + \frac 12 \Ex_{\Cbf_{j+1}} (\theta V ; \theta V )
    \right).
\end{equation}
Let $V_{j+1}'=\Ex_{\Cbf_{j+1}}\theta V_j'$ with initial condition $V_0'=V_0$.
Since $P$ and $W$ are quadratic in $V$, it would be sufficient to
solve
\begin{equation}
\label{e:Wfind2}
    W_{j+1}(V_{j+1}' ) \approx
    \Irr\left(\Ex_{\Cbf_{j+1}} \theta W_j(V_j')
    + \frac 12 \Ex_{\Cbf_{j+1}} (\theta V_{j}'; \theta V_j')
    \right),
\end{equation}
instead of \eqref{e:Wfind1}.
Thus we are led to the problem of showing that $W$ as defined in
\eqref{e:WLTF-bis} satisfies
\eqref{e:Wfind2}.

Starting with $j=0$, for which $W_0=0$,
we set
\begin{equation}
    W_1(V_1')
     = \frac 12 \Irr  \, \Ex_{\Cbf_1} (\theta V_{0}'; \theta V_0').
\end{equation}
For $j=1$, this leads to
\begin{align}
    W_2 (V_2')
    &\approx
    \frac 12 \Irr \left(\Ex_{\Cbf_2} \theta \Irr \,
    \Ex_{\Cbf_1} (\theta V_{0}'; \theta V_0') +
    \Ex_{\Cbf_2} (\theta V_{1}'; \theta V_1') \right)
    \nnb & \approx
    \frac 12 \Irr \left(\Ex_{\Cbf_2} \theta
    \Ex_{\Cbf_1} (\theta V_{0}'; \theta V_0') +
    \Ex_{\Cbf_2} (\theta V_{1}'; \theta V_1') \right)
    ,
\end{align}
where in the second line we used
\eqref{e:IrrRel}.
But by definition,
\begin{equation}
\label{e:Eplus}
    \Ex_{\Cbf_2}\theta \Ex_{\Cbf_1} (\theta V_{0}'; \theta V_0'(\Lambda))
     +
    \Ex_{\Cbf_2} (\theta V_{1}'; \theta V_1')
    =
    \Ex_{\Cbf_1+\Cbf_2}(\theta V_{0}';\theta V_0'),
\end{equation}
and hence
\begin{align}
    W_2 (V_2')
    &\approx
    \frac 12 \Irr \,
    \Ex_{\Cbf_1+\Cbf_2}(\theta V_{0},\theta V_{0}).
\end{align}
Iteration then leads to the stable form
\begin{equation}
\label{e:Wstab}
    W_j (V_j')
    =
    \frac 12 \Irr \,
    \Ex_{\wbf_j}(\theta V_{0},\theta V_0)
    \quad \quad \text{with} \quad \quad \wbf_j=\sum_{i=1}^j \Cbf_i.
\end{equation}
By \eqref{e:Wstab} and \eqref{e:EFABC},
\begin{equation}
\label{e:Wdef}
    W_j (V_j')
    =
    \frac 12 \Irr \,
    F_{\wbf_j}( V_{j}', V_j').
\end{equation}

In the above, $\Rel$ is applied to
$F_{\wbf_j}(V(\Lambda),V(\Lambda)) = \sum_{x \in \Lambda}
F_{\wbf_j}(V_x,V(\Lambda))$.
Naively, we wish to define $\Rel=\LT_\Lambda$, where $\LT$ is
the localisation operator of \cite[Definition~\ref{loc-def:LTXYsym}]{BS-rg-loc}.
A difficulty
with this is that $\Lambda$ is not a coordinate patch in the sense
used in \cite{BS-rg-loc}, so $\LT_\Lambda$ is not defined.
This difficulty is easily overcome
as, inspired by
\cite[Proposition~\ref{loc-prop:LTsum}]{BS-rg-loc}, we can  use
the well-defined quantity $\sum_{x\in \Lambda} \LT_x F_{\wbf_j}(V_x,V(\Lambda))$
instead of the ill-defined $\LT_\Lambda F_{\wbf_j}(V(\Lambda),V(\Lambda))$.
Thus we are led to define
\begin{equation}
\label{e:Reldef}
    \Rel \; F_{\wbf_j}(V(\Lambda),V(\Lambda))
    =
    \sum_{x\in \Lambda} \LT_x  F_{\wbf_j} (V_x,V(\Lambda)).
\end{equation}
In our application, it is shown in Lemma~\ref{lem:EV} below that
$\Ex_{\Cbf} \theta $ maps the range of $\LT$ into itself,
and our assumption \eqref{e:IrrRel} is
then a consequence of \cite[\eqref{loc-e:oLT4b}]{BS-rg-loc}.
Finally, we observe that \eqref{e:Wdef} is consistent with
\eqref{e:WLTF-bis}, and this completes the proof.
\end{proof}

We close this discussion with two further comments concerning $W_j$.
First, although $e^{-V}(1+W)$ and $e^{-V+W}$ are equivalent as
formal power series up to a third order error, they are by no means
equivalent for the expectation.
To illustrate this point with a single-variable example, if $V=\phi^4$ and
$W=\phi^6$, then $e^{-V}(1+W)$ is an integrable function of $\phi$,
but $e^{-V+W}$ is certainly not.  We keep $W$ out of the exponent
in $I$ for reasons related to this phenomenon.

Second, in our applications we use a covariance decomposition with
the finite-range property that
$\wbf_{j;x,y}=0$ if $|x-y|>\half L^{j}$, for some $L>1$.
This is discussed in detail in Section~\ref{sec:Cdecomp} below.
With such a decomposition,
although by definition it appears that $W_j(V,x)$ depends on
$V(\Lambda)$ and hence on the fields at all points in space,
it in fact depends only on $V_y$ with
$|x-y| \leq \half L^{j}$.

\section{Setup and definitions}
\label{sec:main-results-focc}

Now we adapt the discussion of Section~\ref{sec:WPjobs}
to the particular setting of
the supersymmetric field theory representing the
 $4$-dimensional weakly self-avoiding walk,
and make precise definitions of the objects of study, including $\Vpt$.
The minor modifications required
to study the $n$-component $|\varphi|^4$ spin model instead of the weakly self-avoiding walk
are discussed in \cite{BBS-phi4-log}.

\subsection{Fields and observables}
\label{sec:polynomials}

Let $d \ge 4$ and let $\Lambda = \Zd/L^N\Z$ denote the discrete $d$-dimensional
torus of side $L^N$, with $L>1$ fixed, and ultimately with $N\to \infty$.
The field theory we consider consists of a complex boson field $\phi :
\Lambda \to \C$ with its complex conjugate $\bar\phi$, and a pair of
conjugate fermion fields $\psi,\bar\psi$.
The fermion field is given in terms of the
1-forms $d\phi_x$ by $\psi_x = \frac{1}{\sqrt{2\pi i}} d\phi_x$ and
$\bar\psi_x = \frac{1}{\sqrt{2\pi i}} d\bar\phi_x$, where we fix some
square root of $2\pi i$.  This is the supersymmetric choice discussed
in more detail in
\cite[Sections~\ref{norm-sec:df}--\ref{norm-sec:supersymmetry}]{BS-rg-norm}.

In addition, we allow an optional \emph{constant}
complex observable boson field $\sigma \in \C$ with its complex
conjugate $\bar\sigma$.  The observable field is used
in the analysis of the two-point function
of the weakly self-avoiding walk in \cite{BBS-saw4},
and in the more extensive analysis of correlation
functions presented in \cite{ST-phi4}.
Readers only interested in \emph{bulk} quantities, such as the
susceptibility of the weakly self-avoiding walk, may skip any discussion
of observables, or set $\sigma=0$.

For the analysis of the two-point function,
two particular points $\pp,\qq \in \Lambda$ are fixed.
We then work with an algebra $\Ncal$
which is defined in terms of a direct sum decomposition
\begin{equation}
    \Ncal = \Ncal^\varnothing \oplus \Ncal^a \oplus \Ncal^b \oplus \Ncal^{ab}.
\end{equation}
The algebra $\Ncal^\varnothing$ describes the bulk. Its elements
are given by finite linear
combinations of products of fermion fields with coefficients that are
functions of the boson fields.
The algebras $\Ncal^a$, $\Ncal^b$, $\Ncal^{ab}$ account for contributions
due to observables. Their elements
are respectively given by elements of $\Ncal^\varnothing$
multiplied by $\sigma$, by $\bar\sigma$, and by $\sigma\bar\sigma$.
For example, $\phi_x \bar\phi_y \psi_x \bar\psi_x \in
\Ncal^\varnothing$, and $\sigma \bar\phi_x \in \Ncal^a$.
Thus $F \in \Ncal$ has the expansion
\begin{equation}
    \label{e:Fcomponents}
    F
    =
    F_{\varnothing} +
    F_{\pp}\sigma + F_{\qq}\sigmab + F_{\pp \qq} \sigma \sigmab
\end{equation}
with components $F_{\varnothing}, F_{\pp}, F_{\qq}, F_{\pp \qq} \in
\Ncal_{\varnothing}$.  There are canonical projections $\pi_\alpha:
\Ncal \to \Ncal^\alpha$ for $\alpha \in \{\varnothing, a, b, ab\}$.
We use the abbreviation $\pi_*=1-\pi_\varnothing =
\pi_a+\pi_b+\pi_{ab}$.   The algebra $\Ncal$ is also discussed around
\cite[\eqref{loc-e:1Ncaldecomp}]{BS-rg-loc} (there $\Ncal$ is written
$\Ncal/\Ical$ but to simplify the notation we write $\Ncal$ here
instead).

\subsection{Specification of \texorpdfstring{$\LT$}{Loc}}
\label{sec:loc-specs}

As motivated in Section~\ref{sec:WPjobs},
to apply the renormalisation group method,
we require an appropriate projection from $\Ncal$ onto
a finite-dimensional vector space $\Vcal$ of local polynomials in the fields.
This projection is the operator $\LT_X$ defined and discussed in \cite{BS-rg-loc}.
In the absence of observables,
for any set $X \subset \Lambda$, the localisation operator $\LT_X : \Ncal \to \Vcal$ of
\cite[Definition~\ref{loc-def:LTXYsym}]{BS-rg-loc} is simply given by
\begin{align}
    \label{e:LTdef2noob}
    &
    \LT_{X}F =
    \LTsym_{X}^\varnothing F_{\varnothing}
    ,
\end{align}
with $\LTsym_X^\varnothing$ specified below.
In the presence of observables, $\LT_X$
is defined in a graded fashion by
\begin{align}
    \label{e:LTdef2}
    &
    \LT_{X}F =
    \LTsym_{X}^\varnothing F_{\varnothing} +
    \sigma \LTsym_{X\cap \{a \}}^a F_{\pp} +
    \bar\sigma \LTsym_{X\cap \{b \}}^b F_{\qq} +
    \sigma \bar\sigma\LTsym_{X\cap \{a,b \}}^{ab} F_{\pp \qq}
    .
\end{align}
The definition of each $\LTsym^\alpha$ requires: (i) specification of the scaling
(or ``engineering'') \emph{dimensions} of the fields, (ii) choice of a
maximal monomial dimension $d_+=d_+(\alpha)$ for each component of
the range $\Vcal=\Vcal^\varnothing + \Vcal^a + \Vcal^b + \Vcal^{ab}$ of $\LT$,
and (iii) choice
of covariant field polynomials $\hat P$ which form bases for the vector
spaces $\Vcal^\alpha$ (see \cite[Definition~\ref{loc-def:Vcal}]{BS-rg-loc}).

The dimensions of the boson and fermion fields are given by
\begin{equation}
\label{e:phidim}
    [\phi]=[\bar\phi]=[\psi]=[\bar\psi]=
    \textstyle{\frac{d-2}{2}} =1.
\end{equation}
By definition, the dimension of a monomial
$\nabla^\alpha \varphi$ is
equal to $|\alpha|_1+[\phi]$, where $\alpha$ is a multi-index and
$\varphi$ may be any of $\phi,\bar\phi,\psi,\bar\psi$, and the dimension of a product
of such monomials is the sum of the dimensions of the factors in the product.

For the restriction $\LTsym^\varnothing$ of $\LT$ to
$\Ncal^\varnothing$, we take $d_+=d=4$, the spatial dimension.  A
natural way to choose the polynomials $\hat P$ and the space $\Vcal$
they span is given in \cite[\eqref{loc-e:Pm-def}]{BS-rg-loc}.  For
$\LTsym^\varnothing$, we apply the choice given in
\cite[\eqref{loc-e:Pm-def}]{BS-rg-loc} for all monomials in $\Mcal_+$
with maximal dimension $d_+=d=4$, with one exception.  The exception
involves monomials containing a factor $\nabla^e\nabla^{e}\varphi$,
where $\varphi$ may be any of $\phi,\bar\phi,\psi,\bar\psi$.  For
these, we use the choice described in
\cite[Example~\ref{loc-ex:Pm-nonunique}]{BS-rg-loc}, namely we define
$\hat P$ by replacing $\nabla^e\nabla^e\varphi$ by
$\nabla^{-e}\nabla^{e}\varphi$.
The set $\Vcal$ then has the Euclidean invariance property specified
in \cite[Proposition~\ref{loc-prop:EVcal}]{BS-rg-loc}.

In the presence of observables,
the specification of $\LTsym^a$, $\LTsym^b$ and $\LTsym^{ab}$ depends
on the scale $j$, and in particular depends on whether $j$ is above or
below the \emph{coalescence scale} $j_{\pp \qq}$ defined in terms of
the two points $\pp,\qq \in \Lambda$ by
\begin{equation}
   \label{e:Phi-def-jc}
    j_{\pp \qq}
    =
    \big\lfloor
   \log_{L} (2 |\pp - \qq|)
   \big\rfloor
   .
\end{equation}
We assume that $\pi_{ab}V_j=0$ for $j<j_{ab}$, i.e., that $V_j$ cannot
have a $\sigma\bar\sigma$ term before the coalescence scale is
reached.  For $\LTsym^{ab}$ we take $d_+=0$.  When $\LT$ acts at scale
$k$, for $\LTsym^a$ and $\LTsym^b$ we take
$d_+=[\phi]=\frac{d-2}{2}=1$ if $k<j_{\pp\qq}$, and $d_+=0$ for $k \ge
j_{\pp\qq}$.  This choice keeps $\sigma \bar\phi$ in the range of
$\LT$ below coalescence, but not at or above coalescence.
The above phrase ``$\LT$ acts at scale $k$'' means that $\LT$ produces
a scale $k$ object.  For example, $\Vpt$ is a scale $j+1$ object, so
the $\LT$ occurring in $P$ of \eqref{e:beta-function2-H} is considered
to act on scale $j+1$.  Thus the change in specification of $\LT$
occurs for the first time in the formula for the scale $j_{ab}$
version of $\Vpt$.

Moreover, when restricted to $\pi_* \Ncal$, according to
\eqref{e:LTdef2}, $\LT_X$ is the zero operator when $X \cap
\{a,b\}=\varnothing$.  By definition, the map $\LTsym_{X \cap
\{a\}}^{a}$ is zero if $a \not\in X$, and if $a\in X$ it projects onto
the vector space spanned by
$\{1,\phi_a,\bar\phi_a,\psi_a,\bar\psi_a\}$ for $j<j_{\pp\qq}$, and by
$\{1\}$ for $j \ge j_{\pp\qq}$.  A similar statement holds for
$\LTsym_{X \cap \{b\}}^{b}$, whereas the range of $\LTsym_{X \cap
\{a,b\}}^{ab}$ is the union of the ranges of $\LTsym_{X \cap
\{a\}}^{a}$ and $\LTsym_{X \cap \{b\}}^{b}$.  As discussed in
Section~\ref{sec:obssym}, in our application symmetry considerations
reduce the range of $\LT_X$ to the spans of $\{\1_a
\sigma\bar\phi_a\}$, $\{\1_b \bar\sigma \phi_b\}$, and
$\{\1_a\sigma\bar\sigma, \1_b\sigma\bar\sigma\}$, on $\Ncal^a$,
$\Ncal^b$ and $\Ncal^{ab}$, respectively; in fact $\LT_X$ reduces to
the zero operator on $\Ncal^a$, $\Ncal^b$, for $j \ge j_{\pp\qq}$.

This completes the specification of the operator $\LT$.

\subsection{Local polynomials}

The range $\Vcal$ of $\LT$ consists of local polynomials in the fields.
In this paper, we only encounter the subspace $\Qcal \subset \Vcal$
of local polynomials defined as follows.
To define this subspace, we first let $\units$ denote
the set of $2d$ nearest neighbours of the origin in $\Lambda$,
and, for $e \in \units$, define the finite difference
operator $\nabla^e \phi_x = \phi_{x+e}-\phi_x$.  We also set
$\Delta
    =
    -\frac{1}{2}\sum_{e \in \units}\nabla^{-e} \nabla^{e}$.
Then we define the 2-forms
\begin{align}
    \tau_x &= \phi_x \bar\phi_x + \psi_x \bar\psi_x,
\\
\label{e:taunabnabdef}
   \tau_{\nabla \nabla,x}  &=
   \frac 12
   \sum_{e \in \units}
   \left(
   (\nabla^e \phi)_x (\nabla^e \bar\phi)_x +
   (\nabla^e \psi)_x (\nabla^e \bar\psi)_x
   \right)
,
\\
 \label{e:addDelta}
      \tau_{\Delta,x} &= \frac 12 \left(
    (-\Delta \phi)_{x} \bar{\phi}_{y} +
    \phi_{x} (-\Delta \bar{\phi})_{y} +
    (-\Delta \psi)_{x} \bar{\psi}_{y} +
    \psi_{x} (-\Delta \bar{\psi})_{y}
    \right)
    .
\end{align}
The subspace $\Qcal \subset \Vcal$ is then defined to consist of elements
\begin{align}
  \label{e:Vterms}
    &
    V
    =
    g \tau^{2} + \nu \tau
    + y \tau_{\nabla \nabla} + z \tau_{\Delta}
    +
    \lambda_{\pp} \,\sigmaa \bar{\phi}
    +
    \lambda_{\qq} \,\sigmab \phi
    +
    q_{\pp \qq}\sigmaa \sigmab
    ,
\end{align}
where
\begin{align}
\label{e:lambda-defs}
&
    \lambda_{\pp}
    =
    - \lambdaa \,\1_{\pp},
\quad\quad
    \lambda_{\qq}
    =
    - \lambdab \,\1_{\qq},
\quad\quad
    q_{\pp\qq}
    =
    -\frac{1}{2} (\qa\1_{\pp} + \qb\1_{\qq})
    ,
\end{align}
$g,\nu,y,z,\lambdaa,\lambdab,\qa,\qb \in \C$, and the indicator
functions are defined by the Kronecker delta $\1_{a,x}=\delta_{a,x}$.

The observable terms $\lambda_{\pp} \,\sigmaa \bar{\phi} +
\lambda_{\qq} \,\sigmab \phi + q_{\pp \qq}\sigmaa \sigmab$ are
discussed in further detail in Section~\ref{sec:obssym} below.  For
the bulk, the following proposition shows that $\Qcal$ arises as a
supersymmetric subspace of $\Vcal$.  To avoid a digression from the
main line of discussion, the definition of supersymmetry is deferred
to Section~\ref{sec:supersymmetry}, where the proposition is also
proved.

\begin{prop}\label{prop:Qcal}
For the bulk, $\pi_\varnothing\Qcal$ is the subspace of
$\pi_\varnothing\Vcal$ consisting of supersymmetric local
polynomials that are of even degree as forms and without constant
term.
\end{prop}

The fact that constants are not needed in $\Qcal$ is actually
a consequence of supersymmetry
(despite the fact that constants \emph{are} supersymmetric).
This is discussed in Section~\ref{sec:supersymmetry}.

\subsection{Finite-range covariance decomposition}
\label{sec:frcd}

Our analysis involves approximation of $\Zd$ by a torus
$\Lambda = \Zd/L^N\Zd$ of side length $L^N$,
and for this reason we are interested in decompositions of
the covariances
$[\Delta_\Zd + m^2]^{-1}$ and $[-\Delta_\Lambda +m^2]^{-1}$
as operators on $\Zd$ and $\Lambda$, respectively.
For $\Zd$, this Green function exists for $d>2$ for all
$m^2 \ge 0$, but for $\Lambda$ we must take $m^2>0$.
For $\Zd$, in Section~\ref{sec:Cdecomp} we follow \cite{Baue13a}
to define a sequence
$(C_j)_{1 \le j < \infty}$ (depending on $m^2 \ge 0$)
of positive definite covariances
$C_j = (C_{j;x,y})_{x,y\in\Z^d}$
such that
\begin{equation}
\lbeq{ZdCj}
    [\Delta_\Zd + m^2]^{-1} = \sum_{j=1}^\infty C_j
    \quad
    \quad
    (m^2 \ge 0).
\end{equation}
The covariances $C_j$ are
Euclidean invariant, i.e., $C_{j;Ex,Ey} = C_{j;x,y}$
for any lattice automorphism $E: \Z^d \to \Z^d$ (see Section~\ref{sec:supersymmetry}),
and have the \emph{finite-range} property
\begin{equation}
\label{e:frp}
      C_{j;x,y} = 0 \quad \text{if \; $|x-y| \geq \frac{1}{2} L^j$}.
\end{equation}
For $j<N$, the covariances $C_j$ can therefore be identified with
covariances on $\Lambda$, and we use both interpretations.
There is also a covariance $C_{N,N}$ on $\Lambda$ such that
\begin{equation}
\lbeq{NCj}
    [-\Delta_\Lambda + m^2]^{-1} = \sum_{j=1}^{N-1} C_j + C_{N,N}
    \quad
    \quad
    (m^2 > 0).
\end{equation}
Thus the finite volume decomposition agrees with the infinite volume
decomposition except for the last term in the finite volume decomposition.
The special covariance $C_{N,N}$ plays only a minor role in this paper.
For $j \le N$, let
\begin{equation}
\label{e:wdef}
    w_j = \sum_{i=1}^j C_i.
\end{equation}

\subsection{Definition of \texorpdfstring{$\Vpt$}{Vpt}}
\label{sec:Vpt-def}

The finite range decomposition \eqref{e:NCj} is associated
to a natural notion of
\emph{scale}, indexed by $j=0,\ldots, N$.
In our application in \cite{BBS-saw4-log,BBS-saw4},
we are led to consider a family of polynomials $V_j \in \Qcalnabla$ indexed
by the scale.  Our goal here is to describe how, in the second order \emph{perturbative} approximation, these polynomials
evolve as a function of the scale, via the flow of their coefficients,
or \emph{coupling constants}.

Given a positive-definite
matrix $C$ whose rows and columns are indexed by $\Lambda$,
we define the \emph{Laplacian}
(see \cite[\eqref{norm-e:Lapss}]{BS-rg-norm})
\begin{equation}
\label{e:LapC}
    \Lcal_C = \frac 12 \Delta_{C}
=
    \sum_{u,v \in \Lambda}
    C_{u,v}
    \left(
    \frac{\partial}{\partial \phi_{u}}
    \frac{\partial}{\partial \bar\phi_{v}}
    +
    \frac{\partial}{\partial \psi_{u}}
    \frac{\partial}{\partial \bar\psi_{v}}
    \right).
\end{equation}
The Laplacian is intimately related to Gaussian integration.
To explain this, suppose we are
given an additional boson field $\xi,\bar\xi$ and
an additional fermion field $\eta, \bar\eta$, with
$\eta = \frac{1}{\sqrt{2\pi i}}d\xi$,
$\bar\eta = \frac{1}{\sqrt{2\pi i}}d\bar\xi$,
and consider the ``doubled'' algebra $\Ncal(\Lambdabold\sqcup
\Lambdabold')$ containing the original
fields and also these
additional fields.
We define a
map $\theta : \Ncal(\Lambdabold) \to \Ncal(\Lambdabold\sqcup
\Lambdabold')$ by making the replacement
in an element of $\Ncal$ of $\phi$ by $\phi+\xi$,
$\bar\phi$ by $\bar\phi+\bar\xi$, $\psi$ by $\psi+\eta$, and
$\bar\psi$ by $\bar\psi+\bar\xi$.
According to
\cite[Proposition~\ref{norm-prop:conv}]{BS-rg-norm}, for a polynomial $A$
in the fields, the Gaussian super-expectation with covariance $C$ can
be evaluated using the Laplacian operator
via
\begin{equation}
\label{e:EWick}
    \Ex_C \theta A
     = e^{\Lcal_C} A,
\end{equation}
where the fields $\xi,\bar\xi,\eta,\bar\eta$ are integrated out by $\Ex_C$,
with $\phi, \bar\phi, \psi, \bar\psi$ kept fixed, and where $e^{\Lcal_C}$
is defined by its power series.

For polynomials $V',V''$ in the fields, we define
\begin{equation}
\label{e:FCAB}
    F_{C}(V',V'')
    = e^{\Lcal_C}
    \big(e^{-\Lcal_C}V'\big)
    \big(e^{-\Lcal_C}V'' \big) - V'V'',
\end{equation}
By definition, $F_{C} (V',V'')$ is symmetric and bilinear in $V'$ and $V''$.
The map $e^{-\Lcal_C}$ is equivalent
to \emph{Wick ordering} with covariance $C$ \cite{GJ87},
i.e., $e^{-\Lcal_C}A = \, :\!\! A \!\!:_C$.  In
this notation, we could write $F_C$ as a truncated expectation
\begin{equation}
    F_{C}(V',V'')
    =
    \Ex_{C}\theta(:\!\! V' \!\!:_C \, ; :\!\! V'' \!\!:_C),
\end{equation}
but we will keep our expressions in terms of $F_C$.

To handle observables correctly, we also define
\begin{equation}
\label{e:Fpi}
    F_{\pi ,C}(V',V'')
    =
    F_{C}(V',\pi_\varnothing V'')
    + F_{C}(\pi_* V',V'').
\end{equation}
In particular $F_{\pi ,C}$ is the same as $F_{C}$ in the absence of observables,
but not in their presence.
When observables are present,
if $V'$ is expanded as $V'=\pi_{\varnothing} V' + \pi_{*} V'$, there are
cross-terms
$F_C(\pi_\varnothing V', \pi_* V'') + F(\pi_* V',\pi_\varnothing V'')$.
The polynomial \eqref{e:Fpi} is
obtained from \eqref{e:FCAB} by  replacing these cross-terms by
$2 F_{C} (\pi_{*}V' , \pi_{\varnothing}V'')$.
This unusual bookkeeping accounts correctly for observables (it
plays a role in the flow of the coupling constants
$\lambda,q$ and also in estimates in \cite{BS-rg-IE}).

For $X \subset \Lambda$
we define $W_j(V,X)= \sum_{x \in X} W_j(V,x)$ with
\begin{equation}
\label{e:WLTF}
    W_j(V,x) = \frac 12 (1-\LT_{x}) F_{\pi,w_j}(V_x,V(\Lambda)),
\end{equation}
where $\LT_x$ ($=\LT_{\{x\}}$)
is the operator specified above, $V(\Lambda) = \sum_{x \in  \Lambda}
V_x$ as in \eqref{e:VXdef}, and $w_j$ is given by \refeq{wdef}.
Let
\begin{equation}
\label{e:PdefF}
    P(X)=P_j(V,X) = \sum_{x \in X}
    \LT_x \left(
    e^{\Lcal_{j+1}} W_j(V,x)
    + \frac 12
    F_{\pi,C_{j+1}}
    (e^{\Lcal_{j+1}} V_x,e^{\Lcal_{j+1}} V(\Lambda))
    \right),
\end{equation}
where here and throughout the rest of the paper we write
$\Lcal_k = \Lcal_{C_k}$.
Finally, given $V$, we define
\begin{equation}
\label{e:Vptdef}
    \Vpt(V,X)
    =
    e^{\Lcal_{j+1}} V(X)
    -
    P_j(V,X),
\end{equation}
where we suppress the dependence of $\Vpt$ on $j$ in its notation.
The subscript ``pt'' stands for ``perturbation theory''---a
reference to the formal power series calculations discussed
in Section~\ref{sec:WPjobs} that lead to
its definition.
Given $V \in \Vcal$, the polynomial $\Vpt$ also lies in $\Vcal$ by definition.
The polynomial $\Vpt$ is the updated version of $V$ as we move from
scale $j$ to scale $j+1$ via integration of the fluctuation fields
with covariance $C_{j+1}$.

\begin{rk} \label{rk:LocN}
Recall from \cite[\eqref{norm-e:NXdef}]{BS-rg-norm}
that, for $X \subset \Lambda$,
$\Ncal(X)$ consists of those elements of $\Ncal$ which
depend on the fields only at points in $X$ (for this purpose,
we regard the external field
$\sigma$ as located at $\pp$ and $\bar\sigma$ as located
at $\qq$).
A detail needed in the above concerns the
$\Ncal_{X}$ hypothesis in
\cite[Definition~\ref{loc-def:LTsym}]{BS-rg-loc}, which
requires that we avoid applying $\LT$ to elements of $\Ncal(X)$ when
$X$ ``wraps around'' the torus.  We are apparently applying $\LT_x$
in \refeq{WLTF}--\refeq{PdefF}
to field polynomials supported on the entire torus $\Lambda$.
However, the finite-range property \refeq{frp} ensures that the $\Ncal_X$
hypothesis is satisfied for scales $j +1 < N$, so that $\LT$
and $\Vpt$ are well-defined.
For the moment, we do not define $\Vpt$ when $j+1=N$, but we revisit
this in Definition~\ref{def:VptZd} below.
\end{rk}

\subsection{Further definitions}

To prepare for our statement of the explicit computation of $\Vpt$,
some definitions are needed.  The following definitions are all
in terms of the infinite volume decomposition $(C_j)$ of \refeq{ZdCj}.
We write $C=C_{j+1}$ and $w=w_j = \sum_{i=1}^j C_i$.
Given $g,\nu \in \R$, let
\begin{equation}
  \eta' = 2C_{0,0},
  \quad
  \nu_+ = \nu + \eta' g  ,
  \quad
  w_+=w+C,
  \lbeq{nuplusdef}
\end{equation}
and, given a function $f=f(\nu,w)$, let
\begin{equation}
  \label{e:delta-def}
  \delta[f (\nu ,w)]
  =
  f (\nu_+ ,w_{+}) -  f (\nu  ,w )
  .
\end{equation}
For a function $q_{0,x}$ of $x \in \Zd$,
we also define
\begin{equation} \label{e:wndef}
  (\nabla q)^2 = \frac 12 \sum_{e \in \units}(\nabla^e q)^2, \quad
  q^{(n)} = \sum_{x\in \Zd} q_{0,x}^n,\quad
  q^{(**)} = \sum_{x \in \Zd} x_{1}^{2} q_{0,x}.
\end{equation}
All of the functions $q_{0,x}$ that we use are combinations of $w$ that are invariant under lattice rotations,
so that $x_1^2$ can be replaced by $x_i^2$ for any $i=1,\dots, d$ in \eqref{e:wndef}.
We then define
\begin{alignat}{2}
  \lbeq{betadef}
  \beta &= 8 \delta[w^{(2)} ],
  \qquad&\theta  &= 2 \delta[(w^{3})^{(**)}]
  ,
  \\
  \lbeq{xipidef}
  \xi' &=
  4
  \big(
  \delta[w^{(3)}] - 3w^{(2)}C_{0,0}
  \big)
  + {\textstyle{\frac 14}} \beta   \eta',
  \qquad&\pi' &= 2  \delta[(w\Delta w)^{(1)}]
  ,
  \\
  \label{e:sigzetadef}
  \sigma &= \delta[(w\Delta w)^{(**)}],
  \qquad&\zeta &= \delta[((\nabla w)^2)^{(**)}]
  .
\end{alignat}
The dependence on $j$ in the above quantities has been left implicit.

We define a map $\varphi_{\pt}=\varphi_{\pt,j} : \Qcal \to \Qcal$ as
follows.  Given $V$ defined by coupling constants
$(g,\nu,z,y,\lambdaa,\lambdab,\qa,\qb)$, the polynomial
$\varphi_\pt(V)$ has bulk coupling constants
\begin{align}
  \gpt
  &
  =
  g
  - \beta g^{2}   - 4g \delta[\nu w^{(1)}]
  ,
\label{e:gpt2a}
  \\
  \nu_\pt
  &=
  \nu +  \eta' (g + 4g\nu w^{(1)})
  -
  \xi' g^{2}
  -
  {\textstyle{\frac 14}} \beta
  g \nu  - \pi' g(z+y)
  - \delta[\nu^{2} w^{(1)}]
  ,
\label{e:nupta}
  \\
  y_\pt
  &=
  y +
  \sigma gz
  - \zeta gy
  - g \delta [\nu (w^2)^{(**)}],
\label{e:ypta}
  \\
  z_\pt
  &=
  z
  -
  \theta g^{2}
  - \textstyle{\frac{1}{2}} \delta[\nu^{2} w^{(**)}]
  - 2 z \delta[\nu w^{(1)}] - (y_\pt -y)
  .
  \label{e:zpta}
  \intertext{The observable coupling constants
  of $\varphi_\pt(V)$, with $(\lambdapt,\lambda)$
      denoting either $(\lambdaapt,\lambdaa)$ or $(\lambdabpt,\lambdab)$ and
      analogously for $(\qpt,q)$, are given by}
  \lambdapt
  &
  =
  \begin{cases}
  (1 - \delta[\nu w^{(1)}])\lambda & (j+1 < j_{\pp\qq})
  \\
  \lambda & (j+1 \ge j_{\pp\qq}),
  \end{cases}
\label{e:lambdapt2}
  \\
  \qpt
  &
  =
  q
  +
  \lambdaa\lambdab \,C_{\pp\qq}
\label{e:qpt2}
  .
\end{align}

\section{Main results}
\label{sec:mr}

We now present our main results, valid for $d=4$.  In
Section~\ref{sec:fcc}, we give the result of explicit computation of
$\Vpt$ of \refeq{Vptdef}.  The form of $\Vpt$ can be simplified by a
change of variables, and we discuss this transformation and its
properties in Section~\ref{sec:ourflow}.  As explained in
\cite{BBS-saw4-log}, the transformed flow equations for the coupling
constants form part of an \emph{infinite-dimensional} dynamical system
which incorporates non-perturbative aspects in conjunction with the
perturbative flow.  This dynamical system can be analysed using the
results of \cite{BBS-rg-flow}, which have been designed expressly for
this purpose.  To apply the results of \cite{BBS-rg-flow}, certain
hypotheses must be verified, and the results of
Section~\ref{sec:ourflow} also prepare for this verification.

\subsection{Flow of coupling constants}
\label{sec:fcc}

The following proposition shows that for $j+1 <N$, if $V \in
\Qcalnabla$ then $\Vpt \in \Qcalnabla$, and gives the
\emph{renormalised coupling constants} $(\gpt,\nupt , \ypt, \zpt,
\lambdaapt, \lambdabpt, \qapt, \qbpt)$ as functions of the coupling
constants $(g,\nu ,y,z, \lambdaa, \lambdab, \qa, \qb)$ of $V$ and of
the covariances $C=C_{j+1}$ and $w=w_j = \sum_{i=1}^j C_i$.

\begin{prop}
\label{prop:Vptg}
Let $d = 4$ and $0 \le j < N-1$.
If $V \in \Qcalnabla$ then $\Vpt \in \Qcalnabla$ and
\begin{equation}
    V_{\pt,j+1} = \varphi_{\pt,j}.
\end{equation}
In particular, $V_{\pt,j+1}$ is independent of $N$ for $j+1<N$.
\end{prop}

For the observable coupling constant $q$,
in view of our assumption that $\pi_{\pp\qq}V_j =0$ for
$j<j_{\pp\qq}$, and since $C_{j+1;\pp\qq}=0$ if $j+1<j_{\pp\qq}$ by
\refeq{Phi-def-jc} and \refeq{frp}, when $V_{\pt,j+1}$ is constructed
from $V_j$ we also have $q_\pt=0$ for $j+1 <j_{\pp\qq}$, i.e.,
$\pi_{\pp\qq}V_{\pt,j+1} =0$.  This lends consistency across scales to
the assumption that $\pi_{\pp\qq}V_j =0$ for $j<j_{\pp\qq}$. In
fact $C_{j+1;\pp\qq}=0$ if $j+1=j_{\pp\qq}$, but we do not take
advantage of this because it is sensitive to the choice of $\ge$ as
opposed to $>$ in \refeq{frp}.

As mentioned in Remark~\ref{rk:LocN},
the definition of $\Vpt$ breaks down
for $j+1=N$ due to an inability to apply the operator $\LT$ on the
last scale, where the effect of the torus becomes essential.
However, in view of Proposition~\ref{prop:Vptg}, the following
definition of $V_{\pt,N}$ becomes natural.
Moreover, when we prove nonperturbative estimates involving $\Vpt$
in \cite[Proposition~\ref{IE-prop:h}]{BS-rg-IE}, we will see that
this definition of $V_{\pt,N}$ remains effective in implementing
an analogue of Proposition~\ref{prop:I-action}.

\begin{defn}
\label{def:VptZd}
We extend the definition of $V_{\pt,j+1}$ to $j+1= N$
by setting $V_{\pt,N}=\varphi_{\pt,N-1}$.
\end{defn}

The equations \refeq{gpt2a}--\refeq{qpt2} are called
\emph{flow equations}
because they are applied recursively with
$C=C_{j+1}$ and $w=w_j$ updated at each stage of the recursion.
They define a $j$-dependent map $V \mapsto \Vpt$ for $j<N-1$.
The proof of Proposition~\ref{prop:Vptg} is by
explicit calculation of
\eqref{e:Vptdef}.  The calculation is mechanical, so mechanical
that it can be carried out on a computer.
In fact,
we have written a program \cite{BBS-rg-ptsoft}
in the Python programming language
to compute the polynomial $P$ of \eqref{e:PdefF}, and this computer program
leads to the explicit formulas given in Proposition~\ref{prop:Vptg}.
From that perspective, it is possible now to write ``QED''
for Proposition~\ref{prop:Vptg}, but in Section~\ref{sec:Vpt} we
nevertheless present a useful Feynman diagram formalism
and use it to derive the coefficients
\refeq{gpt2a} and \refeq{lambdapt2}--\refeq{qpt2} of $\Vpt$.
The same formalism can be used for
\refeq{nupta}--\refeq{zpta}, but we do not present those details
(several pages of mechanical computations).
In Section~\ref{sec:Vpt}, we also discuss consequences of supersymmetry
for the flow equations.

\subsection{Change of variables and dynamical system}
\label{sec:ourflow}

The observable coupling constants do not appear in the flow of the
bulk coupling constants.  Thus the flow equations
\refeq{gpt2a}--\refeq{qpt2} have a \emph{block triangular} structure;
the flow of the bulk coupling constants is the same whether or not
observables are present, whereas the observable flow does depend on the bulk flow.
 This structure is conceptually
important and general; it persists \emph{non-perturbatively} (see \cite{BBS-saw4}
and also \cite{BS-rg-step}),
and also holds for observables used in the analysis of correlation functions other
than the two-point function \cite{ST-phi4}.

We now discuss a change of variables that simplifies the
bulk flow equations.  In particular, the change of variables
creates a system of equations that is itself triangular to
second order.
Unlike the block triangularity in bulk and observable variables,
this triangularity \emph{in the second-order approximation} of the bulk flow
will be broken by higher-order corrections. Nonetheless, it provides an important
structure in our analysis by enabling the application of \cite{BBS-rg-flow}.

In preparation of the definition of the change of variables,
to counterbalance an exponential decay in $\nu_j$,
we define the rescaled coupling constant
\begin{equation}
\lbeq{munu}
    \mu_j = L^{2j}\nu_j,
\end{equation}
and also define normalised coefficients
\begin{gather}
  \label{e:wbardef1}
  \omega_j = L^2 {\textstyle{\frac 14}} \beta_j,
  \quad
  \gamma_j = L^{2(j+1)}\gamma_j' \quad (\gamma=\eta,\xi,\pi),
  \\
  \label{e:wbardef2}
  \bar{w}_j^{(1)} = L^{-2j}  w_j^{(1)},
  \quad
  \bar{w}_j^{(**)} = L^{-4j}w_j^{(**)}
  .
\end{gather}
The constants in \eqref{e:wbardef1}--\eqref{e:wbardef2} are all uniformly bounded,
as we show in Lemma~\ref{lem:wlims}.
Also, summation by parts on the torus gives
$\sum_{x\in\Lambda} \tau_{\nabla\nabla,x}
  =  \sum_{x\in\Lambda} \tau_{\Delta,x}$,
and hence
\begin{equation}
\label{e:zysbp}
    \zpt   \sum_{x\in\Lambda} \tau_{\Delta,x}
    + \ypt \sum_{x\in\Lambda} \tau_{\nabla\nabla,x}
    =
    (\zpt + \ypt) \sum_{x\in\Lambda} \tau_{\Delta,x}.
\end{equation}
Boundary terms do arise if the sum over $\Lambda$ is replaced by
a sum over a proper subset of $\Lambda$, and in \cite{BS-rg-IE,BS-rg-step}
we work with such smaller sums.  Nevertheless, we are able to
make use of \refeq{zysbp} (our implementation occurs in
\cite[Section~\ref{step-sec:int-by-parts2}]{BS-rg-step}).
This suggests that $\zpt+\ypt$ should be a natural variable, so we define
\begin{equation}
    z^{(0)} = y+z,
    \quad\quad
    \zpt^{(0)}= \ypt+\zpt .
\end{equation}

Taking the above into account, given $V$
we define $\Vpt^{(0)}$ by
\begin{equation}
    \Vpt^{(0)}= \gpt \tau^2 + \mupt L^{-2(j+1)} \tau + \zpt^{(0)} \tau_\Delta
    .
\end{equation}
The above definition is valid for \emph{all} $0 \le j < \infty$, using
the formulas \refeq{gpt2a}--\refeq{zpta} with coefficients computed
from the decomposition $(C_{j})_{1 \le j < \infty}$ of
$[-\Delta_\Zd+m^2]^{-1}$.
In view of \eqref{e:gpt2a}--\eqref{e:zpta},
this leads us to consider the equations:
\begin{align}
  \gpt
  &
  =
  g
  - \beta_j g^{2}   - 4g \delta[\mu \bar w^{(1)}]
  ,
  \label{e:newflow-g}
  \\
  \zpt^{(0)}
  &=
  z^{(0)}
  -
  \theta_j g^{2}
  - \textstyle{\frac{1}{2}} \delta[\mu^{2} \bar w^{(**)}]
  - 2 z^{(0)} \delta[\mu \bar w^{(1)}]
  ,
  \label{e:newflow-z}
  \\
  \mu_\pt
  &=
  L^2 \mu +  \eta_j (g + 4g\mu \bar w^{(1)})
  -
  \xi_j g^{2}
  -
  \omega_j
  g \mu  - \pi_j g z^{(0)}
  - \delta[\mu^{2} \bar w^{(1)}]
  \label{e:newflow-mu}
  ,
\end{align}
and we define a map $\varphi^{(0)}_{\pt}=\varphi^{(0)}_{\pt,j}$ on $\R^3$,
for $1 \le j < \infty$, by
\begin{equation}
\label{e:varphiptdef}
    \varphi^{(0)}_{\pt,j}(g, \mu, z^{(0)}) = (g_\pt, \mu_\pt, z_\pt^{(0)}).
\end{equation}

The four terms involving $\delta$ on the right-hand sides
of \refeq{newflow-g}--\refeq{newflow-mu} can be eliminated by a
change of variables.
To describe the transformed system, we define a map
$\bar\varphi_j$ on $\R^3$, for $1 \le j < \infty$, by
\begin{equation}
    \bar\varphi_j(\gbar_j, \zbar_j, \mubar_j)
    =(\gbar_{j+1}, \zbar_{j+1}, \mubar_{j+1}),
\end{equation}
where
\begin{align}
\lbeq{newflow-gbar}
  \gbar_{j+1}
  &=
  \gbar_j
  -
  \beta_j  \gbar_j^{2}
  ,
  \\
\lbeq{newflow-zbar}
  \zbar_{j+1}
  &=
  \zbar_j
  - \theta_j  \gbar_j^{2}
  ,
  \\
\lbeq{newflow-mubar}
  \mubar_{j+1}
  &=
  L^2 \mubar_j
  + \eta_j \gbar_j
  -
  \xi_j  \gbar_j^{2}
  -
  \omega_j
  \gbar_j  \mubar_j
  - \pi_j  \gbar_j  \zbar_j
  .
\end{align}
The change of variables is defined by a polynomial map $T_j:\R^3 \to \R^3$,
which we write as $T_j (g,z,\mu)= (\gch,\zch,\much)$, where
\begin{align}
  \label{e:gch-def1}
  \gch  & = g +4g\mu \bar w_j^{(1)},
  \\
  \label{e:zch-def1}
  \zch & = z + 2z\mu \bar w_j^{(1)} + \tfrac{1}{2}  \mu^2 \bar w_j^{(**)},
  \\
  \label{e:much-def1}
  \much & = \mu + \mu^2 \bar w_j^{(1)}
  .
\end{align}%
The following proposition, which is proven in Section~\ref{sec:pf-transformation},
gives the properties of the change of variables.
We think that the existence of this change of variables
may express an invariance property of the flow equations
with respect to change of covariance decompositions,
one that we do not fully understand.

Below \refeq{Tcomp} and in the remainder of the paper, $O(A^{-k})$
with $k$ unspecified denotes a quantity that is bounded by
$k$-dependent multiple of $A^{-k}$ for arbitrary $k>0$.

\begin{prop} \label{prop:transformation}
Let $d=4$ and $\bar m^2 >0$.
There exist an open ball $B \subset \R^3$ centred at $0$
(independent of $j\ge 1$ and $m^2 \in [0,\bar m^2]$),
and analytic maps $\rho_{\pt,j} : B \to \R^3$ such that,
with the quadratic polynomials $T_j: \R^3 \to \R^3$
given by \eqref{e:gch-def1}--\eqref{e:much-def1},
\begin{equation}
\lbeq{Tcomp}
  T_{j+1} \circ \varphi_{\pt,j}^{(0)}
  =
  \bar\varphi_j \circ T_j + \rho_{\pt,j} \circ T_j
  ,
\end{equation}
$T_j(V)=V+O(|V|^2)$,
the inverse $T_j^{-1}$ to $T_j$
exists on $B$ and is analytic with $T_j^{-1}(V) = V +O (|V|^2)$, and
 $\rho_{\pt,j}(V) = O((1+m^2L^{2j})^{-k}|V|^3)$.
All constants are uniform in~$j\ge 1$ and $m^2 \in [0,\bar m^2]$.
\end{prop}

Our analysis of the dynamical system arising from
the bulk flow equations \eqref{e:gpt2a}--\eqref{e:zpta} is based
on an application of the main result of \cite{BBS-rg-flow}
to the transformed system $\bar\varphi + \rho_{\pt}$.
The main result of \cite{BBS-rg-flow} requires the verification
of \cite[Assumptions~(A1--A3)]{BBS-rg-flow}.
We first recall the statements of \cite[Assumptions~(A1--A2)]{BBS-rg-flow} in our present context,
which are bounds on the coefficients in \eqref{e:newflow-gbar}--\eqref{e:newflow-mubar}.
These coefficients depend on the mass $m^2$ and decay as $j\to\infty$ if $m^2 > 0$.

This decay is naturally measured in terms of the \emph{mass scale} $j_m$,
defined by
\begin{equation} \label{e:jmdef}
     j_m =
     \begin{cases}
     \lfloor \log_{L^2} m^{-2}\rfloor & (m^2>0)
     \\
     \infty & (m^2=0).
    \end{cases}
\end{equation}
However, \cite[Assumptions~(A1--A3)]{BBS-rg-flow} are stated in a more general context,
involving a quantity $j_\Omega$ which is closely related to $j_m$.
To define $j_\Omega$, we fix $\Omega > 1$, and set
\begin{equation}
  \lbeq{jOmegadef}
  j_\Omega = \inf \{ k \geq 0: |\beta_j| \leq \Omega^{-(j-k)} \|\beta\|_\infty
  \text{ for all $j$} \}
  ,
\end{equation}
with $j_\Omega = \infty$ if the infimum is over the empty set.
In Proposition~\ref{prop:rg-pt-flow} below it is shown that
$j_\Omega = j_m + O(1)$,
uniformly in $m^2 \in (0,\delta]$, with $j_m=j_\Omega=\infty$ if $m^2=0$.

Assumption~(A1) asserts that $\|\beta\|_\infty < \infty$ and that there exists
$c>0$ such that $\beta_j \ge c$ for all but $c^{-1}$ values of $j \le \jm$, while
Assumption~(A2) asserts that each of $\theta_j,\eta_j,\xi_j,\omega_j,\pi_j$
is bounded above in absolute value by $O(\Omega^{-(j-j_\Omega)_+})$ (the coefficient $\zeta_j$
of Assumption~(A2) is zero here).
The result of \cite{BBS-rg-flow} also takes into account
non-perturbative aspects of the flow, which are discussed in \cite{BS-rg-step}.
The following proposition prepares the ground for the application of
\cite{BBS-rg-flow} by verifying that the transformed flow obeys
\cite[Assumptions~(A1--A2)]{BBS-rg-flow}.
We write $\Vch = T(V)$.

\begin{prop} \label{prop:rg-pt-flow}
  Let $d=4$ and $\bar m^2>0$. Each coefficient
  in \refeq{gpt2a}--\refeq{qpt2} and
  in \eqref{e:newflow-gbar}--\eqref{e:newflow-mubar} is a continuous
  function of $m^2 \in [0,\bar m^2]$.
  Fix any $\Omega > 1$.
  For $m^2 \in [0,\delta]$, with $\delta>0$ sufficiently small,
  the map $\bar\varphi$  satisfies
  \cite[Assumptions~(A1--A2)]{BBS-rg-flow}.
  In addition,
  \begin{equation} \label{e:jmjOmega}
    j_\Omega = j_m + O(1)
  \end{equation}
  uniformly in $m^2 \in (0,\delta]$, with $j_m=j_\Omega=\infty$ if $m^2=0$.
\end{prop}

\section{Flow equations and Feynman diagrams}
\label{sec:Vpt}

As mentioned previously, we have written a computer program  \cite{BBS-rg-ptsoft} in
the Python programming language
to compute $\Vpt$, and this program
produces the equations of Proposition~\ref{prop:Vptg}.
In this section, we provide a Feynman diagram formalism, of
independent interest, for
an alternate computation of $\Vpt$.  We use the formalism to derive
the flow equations for $\gpt, \lambdapt, \qpt$ of
\eqref{e:gpt2a} and \eqref{e:lambdapt2}--\eqref{e:qpt2}.
Using this formalism, it is possible also to derive
\eqref{e:nupta}--\refeq{zpta}, but we do not provide those
details.

The polynomial $\Vpt = e^{\Lcal}
V - P$ is defined in \eqref{e:Vptdef}.
In Section~\ref{sec:closed-loops}, we develop the Feynman
diagram approach that we use to calculate
the $\tau^2$ term of $P$,
and compute the term $e^\Lcal V$.
In Section~\ref{sec:supersymmetry}, we
discuss the symmetries of the model and show that they ensure
that $\pi_\varnothing \Vpt$ does not contain any terms
that are not in $\Qcalnabla$, and we prove Proposition~\ref{prop:Qcal}.
Then in Section~\ref{sec:Pcalc} we complete the proof of
\eqref{e:gpt2a} and \eqref{e:lambdapt2}--\eqref{e:qpt2}.
We assume throughout that $d=4$.

\subsection{Feynman diagrams}
\label{sec:closed-loops}

A convenient way to carry out the computation of
$\Vpt$ is via the Feynman diagram notation
introduced in this section.
Given $a,b \in \Lambda$, let
\begin{equation}
    \tau_{ab}
    =
    \phi_{a} \bar{\phi}_{b} + \psi_{a} \bar{\psi}_{b}.
\end{equation}

\begin{lemma}
\label{lem:noloops}
For $a,b \in \Lambda$,
\begin{equation}
    \label{e:no-loops}
    \Lcal_C \tau_{ab}
    =
    0,
\end{equation}
and, for $a_i,b_i \in \Lambda$ and $n \ge 2$,
\begin{equation}
\label{e:contraction}
    \Lcal_C
    \prod_{i=1}^{n} \tau_{a_{i}b_{i}}
    =
    \sum_{i,j:i\not =j}
    C_{b_{i},a_{j}}\tau_{a_{i}b_{j}}
    \prod_{k\not =i,j} \tau_{a_{k}b_{k}}.
\end{equation}
\end{lemma}

\begin{proof}
Equation~\eqref{e:no-loops} follows from \eqref{e:LapC} together with
\begin{equation}
    \left(
    \frac{\partial}{\partial \phi_{x}}
    \frac{\partial}{\partial \bar{\phi}_{y}}
    +
    \frac{\partial}{\partial \psi_{x}}
    \frac{\partial}{\partial \bar{\psi}_{y}}
    \right)
    \tau_{ab}
    \nnb
    =
    \frac{\partial}{\partial \phi_{x}}
    \frac{\partial}{\partial \bar{\phi}_{y}}
    \phi_{a} \bar{\phi}_{b}
    +
    \frac{\partial}{\partial \psi_{x}}
    \frac{\partial}{\partial \bar{\psi}_{y}}
    \psi_{a} \bar{\psi}_{b}
    =
    \delta_{xa} \delta_{yb} - \delta_{xa} \delta_{yb}
    =
    0.
\end{equation}
Also,  taking anti-commutativity into account,
direct calculation gives
\begin{equation}
    \label{e:contract}
    \Lcal_C\,
    \tau_{a_{1}b_{1}}
    \tau_{a_{2}b_{2}}
    =
    C_{b_{1},a_{2}}\tau_{a_{1}b_{2}} +
    C _{b_{2},a_{1}}\tau_{a_{2}b_{1}},
\end{equation}
which is the $n=2$ case of \eqref{e:contraction}.

The general case of \eqref{e:contraction} can then be proved
via induction on $n$, and we just sketch the idea.  First,
we write $\prod_{i=1}^{n} \tau_{a_{i}b_{i}}
=\tau_{a_nb_n} \prod_{i=1}^{n-1} \tau_{a_{i}b_{i}}$.
When $\Lcal_C$ is applied
to the product, there is a contribution of zero when it acts
entirely on the factor $\tau_{a_nb_n}$
and the induction hypothesis can be
applied to evaluate the contribution when $\Lcal_C$ acts entirely
on the factor $\prod_{i=1}^{n-1} \tau_{a_{i}b_{i}}$.  What remains is the
contribution where $\Lcal_C$ acts jointly on both factors, and this can
be seen to give rise to
\eqref{e:contraction}.
\end{proof}

This allows for a very simple calculation of the term
$e^{\Lcal_C} V$ in $\Vpt$, as follows.

\begin{lemma}
\label{lem:EV}
For $V \in \Qcalnabla$,
\begin{align}
\label{e:EV}
    &
    e^{\Lcal_C} V_x
    =
    V_x + 2g C_{x,x}\tau_x .
\end{align}
\end{lemma}

\begin{proof}
Since $V$ is fourth order in the fields, we can expand
$e^{\Lcal_C}$ to second order in $\Lcal_C$ to obtain
\begin{equation}
    e^{\Lcal_C} V
    =
    V + \Lcal_C (g\tau_x^2 + \nu \tau_x + z\tau_{\Delta,x} + y\tau_{\nabla\nabla,x})
    + \frac{1}{2!} \Lcal_C^2 g\tau_x^2.
\end{equation}
In the second term on the right-hand side, it follows from \eqref{e:no-loops}
that only $g\tau_x^2$ yields a nonzero contribution,
and by the $n=2$ case of \refeq{contraction}
this contributes $2gC_{x,x}\tau_x$.
A second application of \eqref{e:no-loops} then shows that
the final term on the right-hand side is zero.
\end{proof}

Lemma~\ref{lem:noloops} shows, in particular, that products of $\tau_{ab}$
remain products of $\tau_{ab}$ under repeated application of the
Laplacian $\Lcal_C$.
In \eqref{e:contract}, we say that $(a_{1},b_{2})$ and $(a_{2},b_{1})$
are \emph{contractions} of $(a_{1},b_{1})$ and $(a_{2},b_{2})$. We
visualise $\tau_{ab}$ as a vertex with an ``in-leg'' labelled $a$ and
an ``out-leg'' labelled $b$:
\begin{center}
\includegraphics[scale = 0.8]{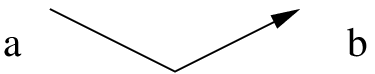}.
\end{center}
Contraction is then the operation of joining an out-leg of a vertex
to an in-leg of a vertex, denoted:
\begin{center}
\includegraphics[scale = 0.8]{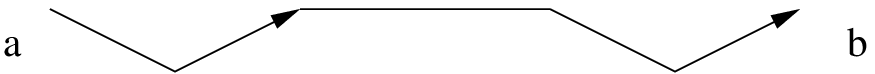}.
\end{center}
Thus we regard \eqref{e:contraction} as the sum over all ways to
contract two of the labelled pairs.  For example, one term that arises
in calculating $\Lcal_C^{2} \prod_{i=1}^{4}
\tau_{a_{i}b_{i}}$ is $C_{a_1,b_3}\tau_{a_1,b_3}C_{a_2,b_4}\tau_{a_2,b_4}$,
which is denoted:
\begin{center}
\includegraphics[scale = 0.8]{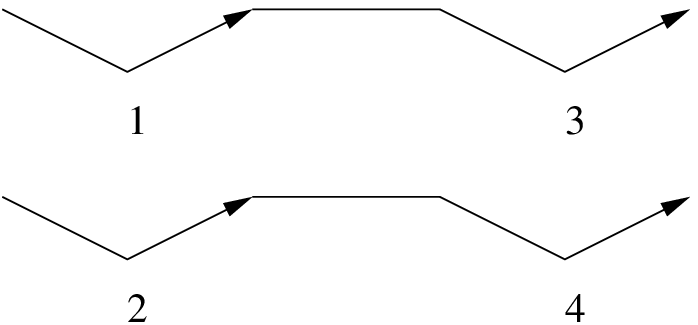}.
\end{center}
This Feynman diagram notation is useful in
Section~\ref{sec:Pcalc}.

\subsection{Symmetries}
\label{sec:supersymmetry}

Next, we discuss the symmetries of the model, and prove
Proposition~\ref{prop:Qcal}.  We first discuss the bulk, and prove in
particular that if $V\in \Qcal$ then $\pi_\varnothing \Vpt \in \Qcal$.
Finally, we discuss the situation for observables.

\subsubsection{Symmetry and the bulk}
\label{sec:bulksym}

\smallskip\noindent
\emph{Supersymmetry.}
Supersymmetry is discussed  in \cite[Section~6]{BIS09}, where
the supersymmetry generator is defined in terms of the exterior derivative
and interior product as
${Q} = d + \ci$.  It is convenient to define $\hat{Q} = (2 \pi i)^{-1/2}Q$.
In our present notation,  $\hat{Q}$ can be written as the
antiderivation on $\Ncal$ defined by
\begin{equation}
\label{e:Qhat}
    \hat Q
    =
    \sum_{x\in \Lambda} \left(
    \psi_{x} \frac{\partial}{\partial \phi_{x}} +
    \bar\psi_{x} \frac{\partial}{\partial \bar\phi_{x}}-
    \phi_{x} \frac{\partial}{\partial \psi_{x}}  +
    \bar\phi_{x} \frac{\partial}{\partial \bar\psi_{x}}
    \right).
\end{equation}
In particular,
\begin{align}
\label{e:Qaction}
   &
   \hat{Q}\phi_{x} = \psi_{x},
   \quad\quad
   \hat{Q}\bar\phi_{x} = \psib_{x},
   \quad \quad
   \hat{Q}\psi_{x} = - \phi_{x},
   \quad \quad
   \hat{Q}\psib_{x} = \bar\phi_{x}
   .
\end{align}
An element $F \in \Ncal$ is said to be \emph{supersymmetric} if $QF=0$.

\smallskip\noindent \emph{Gauge symmetry.}
The \emph{gauge flow} on $\Ncal$ is characterised by
$q \mapsto e^{-2\pi i t}q$ for $q = \phi_{x} ,\psi_{x}$
and
$\bar q \mapsto e^{+2\pi i t}\bar q$ for $\bar q = \bar\phi_{x}, \psib_{x}$, for all
$x \in \Lambda$.
An element $F \in \Ncal$ is said to be
\emph{gauge invariant} if it is invariant under this flow.

\smallskip
As discussed  in \cite[Section~6]{BIS09},
$Q^{2}$ is the generator of the gauge flow, so $F \in \Ncal$ is
gauge invariant if and only if $Q^2 F=0$.  In particular,
supersymmetric elements
are gauge invariant.
Since the boson and fermion fields have the same
dimension,  $Q$ maps $\Vcal$ to itself.
It is straightforward to verify that the monomials in $\pi_\varnothing \Qcal$
are all supersymmetric, hence gauge invariant.

\smallskip
We say that $V\in \Vcal$ is an \emph{even form} if it is a sum of
monomials of even degree in $\psi,\psib$, and we say that $V$ is
\emph{homogeneous of degree $n$} if $V$ lies in the span of monomials
of degree $n$.  For $d= 4$, with fields $\phi ,\bar\phi,\psi,
\bar\psi$ of dimension $[\phi] = 1$, and with $d_+=d=4$, the highest
degree monomials have degree $4$ and can have no spatial derivatives.
Degree $2$ monomials have at most two spatial derivatives. Gauge
invariant monomials in $\phi,\bar\phi,\psi,\bar\psi$ must have even
degree because for every field in the monomial, the conjugate of that
field must also be in the monomial.  The next lemma characterises the
monomials in $\Vcal$ that respect symmetries of the model, and shows
that these are the ones that occur in $\pi_\varnothing\Qcal$.

\begin{lemma}\label{lem:ss}
If $V \in \Vcal$ is even, supersymmetric, and degree $4$, then $V = \alpha
\tau^{2}$ for some $\alpha \in \Cbold$.
If $V \in \Vcal$ is even, homogeneous of degree $2$, and supersymmetric,
then $V$ is a linear combination of $\tau$,
$\tau_{\nabla \nabla}$ and $\tau_{\Delta}$.
\end{lemma}

\begin{proof}
The only gauge invariant, degree four monomials in $\Vcal$ that are even forms
are $(\phi \bar\phi)^{2}$ and $\phi \bar\phi \psi \psib$, because $\psi^{2}=
\psib^{2}=0$. Therefore, for some $\alpha ,\beta \in \Cbold$,
\begin{equation}
   V
   =
   \alpha (\phi \bar\phi)^{2} + \beta \phi \bar\phi \psi \psib
   .
\end{equation}
Recall that $\hat Q$ is an antiderivation.
Since $\psi \psib \hat{Q} (\phi \bar\phi) = 0$ and since $V$ is
supersymmetric,
\begin{align}
   0
   =
   \hat{Q}V
   &
   =
   2\alpha \phi \bar\phi \hat{Q} (\phi \bar\phi)
   +
   \beta \phi \bar\phi \hat{Q} (\psi \psib)
   =
   \phi \bar\phi
   \hat{Q}
   \big(
   2\alpha \phi \bar\phi
   +
   \beta \psi \psib
   \big)
   ,
\end{align}
which by \eqref{e:Qaction} implies that $\beta= 2\alpha $.
Therefore, as required,
\begin{equation}
   V
   =
   \alpha
   \big(
   (\phi \bar{\phi})^{2} + 2 \phi \bar{\phi} \psi
   \psib
   \big)
   =
   \alpha
   \big(
   \phi \bar{\phi} + \psi
   \psib
   \big)^{2}
   =\alpha \tau^{2}
   .
\end{equation}

The monomials in $\Vcal$ which are even, homogeneous of degree $2$,
and gauge invariant are given by
\begin{equation}
   \phi \phib,
\quad
   \sum_{e \in \units}
   \nabla^{e} \phi \nabla^{e} \phib,
\quad
   \phi \Delta \phib +
   (\Delta \phi) \phib ,
\end{equation}
and the same with $\phi$ replaced by $\psi$
(the fact that only Euclidean symmetric monomials occur in $\Vcal$ is
guaranteed by \cite[Proposition~\ref{loc-prop:EVcal}]{BS-rg-loc}.)
If we now impose
supersymmetry, by seeking linear combinations that are annihilated
by $\hat{Q}$, the supersymmetric combination that contains
$\phi \phib$ is $\tau$. Similarly $\tau_{\nabla \nabla}$ and
$\tau_{\Delta}$ are generated by the other two terms.
\end{proof}

\smallskip \noindent
\begin{proof}[Proof of Proposition~\ref{prop:Qcal}.]
  By Lemma~\ref{lem:ss}, the monomials in $\pi_\varnothing\Vcal$ that
  are supersymmetric, of even degree as forms, of even degree in the
  fields, and without constant term, are precisely those in
  $\pi_\varnothing \Qcal$.  Since gauge symmetry eliminates monomials
  that are of odd degree in the fields, this completes the proof.
\end{proof}

\begin{lemma}
\label{lem:ss2}
If $V\in \Qcalnabla$
then $\pi_\varnothing \Vpt \in \Qcalnabla$.
\end{lemma}

\begin{proof}
Since $\pi_\varnothing \Vpt(V) = \Vpt(\pi_\varnothing V)$ (because any
$\sigma$ or $\sigmab$ in $V$ cannot disappear in creation of $\Vpt$),
we can and do assume that $V=\pi_\varnothing V$.  In view of
Lemmas~\ref{lem:EV}--\ref{lem:ss}, it suffices to show that $P$ is
supersymmetric but with no constant term (constants are certainly
supersymmetric).

We begin by showing that $P$ does not contain a nonzero constant term.
Since we are assuming $\pi_*V=0$,
we may replace $F_\pi$ by $F$ in \eqref{e:PdefF}, and also in the
definition of $W$ in \eqref{e:WLTF}.
To see that $F$ contains no constant term,
observe that in \eqref{e:FCAB}, $A,B$ do not contain constant terms,
and therefore, by Lemma~\ref{lem:noloops}, neither do
$e^{-\Lcal}A$ and $e^{-\Lcal} B$.
Hence, again by Lemma~\ref{lem:noloops}, $F$ cannot contain any constant terms.
Therefore, neither does $\LT_x F$.
Similar reasoning shows that the $W$
term in \eqref{e:PdefF} cannot contain a nonzero constant term.
Therefore $P$ does not contain a nonzero constant term.

It remains to show that $P$ is supersymmetric.
Examination of \eqref{e:PdefF} reveals that the supersymmetry of $V$ will
be inherited by $P$ as long as the supersymmetry generator $Q$ commutes
with both $e^{\pm \Lcal}$ and $\LT_x$.
For the former, from \eqref{e:Qhat}, we obtain the commutator formulas
\begin{align}
\label{e:Qcomm}
    \left[
    \frac{\partial }{\partial \phi_u  }
    \frac{\partial }{\partial \bar\phi_v  },
    \hat Q
    \right]
    &=
    -
    \frac{\partial }{\partial \bar\phi_v }
    \frac{\partial }{\partial \psi_u }
    +
    \frac{\partial }{\partial \phi_u }
    \frac{\partial }{\partial \bar\psi_v },
    \\
\label{e:Qcomm2}
    \left[
    \frac{\partial }{\partial \psi_u }
    \frac{\partial }{\partial \bar\psi_v },
    \hat Q
    \right]
    &=
    -
    \frac{\partial }{\partial \bar\psi_v }
    \frac{\partial }{\partial \phi_u }
    +
    \frac{\partial }{\partial \psi_u }
    \frac{\partial }{\partial \bar\phi_v },
\end{align}
and thereby conclude that $\hat Q$ commutes with $\Lcal$ and hence also with
$e^{\pm \Lcal}$.
Finally, the fact that  $Q$ commutes $\LT_x$ is a consequence of
\cite[Proposition~\ref{loc-prop:Qcom}]{BS-rg-loc}.  This completes
the proof.
\end{proof}

\subsubsection{Symmetry and observables}
\label{sec:obssym}

Next, we discuss symmetry of the observables, and the monomials in
$\pi_*\Qcal$.

Recall from \refeq{Fcomponents} that an element $F\in \Ncal$
decomposes as $F= F_{\varnothing} + F_{\pp}\sigma + F_{\qq}\sigmab +
F_{\pp \qq} \sigma \sigmab$, as a consequence of the direct sum
decomposition $\Ncal = \Ncal^\varnothing \oplus \Ncal^a \oplus \Ncal^b
\oplus \Ncal^{ab}$.  The direct sum decomposition of $\Ncal$ induces a
decomposition $\Vcal = \Vcal^{\varnothing} \oplus \Vcal^a \oplus
\Vcal^b \oplus \Vcal^{ab}$.  In particular, each $V \in \Qcal \subset
\Vcal$ is the sum of $ \pi_\varnothing V=g \tau^{2} + \nu \tau + z
\tau_{\Delta}+ y \tau_{\nabla \nabla}$, $\pi_\pp V = \lambda_{\pp}
\,\sigmaa \bar{\phi}$, $\pi_{\qq}V = \lambda_{\qq} \,\sigmab \phi$,
and $\pi_{\pp\qq}V = q_{\pp \qq}\sigmaa \sigmab$.

According to Section~\ref{sec:loc-specs}, the list of monomials in
$\pi_*\Vcal$, i.e., those that contain $\sigma$ and/or $\bar\sigma$,
is as follows.  The monomials containing $\sigma$ but not $\bar\sigma$
are given by $\sigma$ multiplied by any element of
$\{\1_a,\1_a\phi_a,\1_a\bar\phi_a,\1_a\psi_a,\1_a\bar\psi_a\}$ for
$j<j_{\pp\qq}$, and $\sigma$ multiplied by $\{\1_a\}$ for $j \ge
j_{\pp\qq}$.  The monomials containing $\sigmab$ but not $\sigma$
consist of a similar list with $\sigma$ replaced by $\sigmab$ and $a$
replaced by $b$.  The monomials containing $\sigma\sigmab$ are
$\{\1_a\sigma\sigmab,\1_b\sigma\sigmab\}$.

We define the gauge group to act on $\sigma$ and $\sigmab$ via $\sigma
\mapsto e^{-2\pi it}\sigma$ and $\sigmab \mapsto e^{2\pi it}\sigmab$.
If we now demand gauge invariance, and also rule out constants and
forms of odd degree, the remaining monomials in $\pi_*\Vcal$ are
$\{\1_a\sigma\phib_a,\1_b\sigmab \phi_b, \1_a\sigma\sigmab,
\1_b\sigma\sigmab\}$ when $j<j_{\pp\qq}$, and $\{\1_a\sigma\sigmab,
\1_b\sigma\sigmab\}$ when $j\ge j_{\pp\qq}$.

\subsection{Calculation of \texorpdfstring{$P$}{P}}
\label{sec:Pcalc}

It follows from Lemma~\ref{lem:ss2} that
$\pi_\varnothing \Vpt \in \Qcalnabla$, and hence the bulk part
of $\Vpt$ contains only the monomials listed in Lemma~\ref{lem:ss}.
Thus to compute the bulk part of $\Vpt$ it is only necessary to
compute $\gpt, \nupt, \ypt, \zpt$.
In this section, we complete the proof of
\eqref{e:gpt2a} and \eqref{e:lambdapt2}--\eqref{e:qpt2}.
We prove \eqref{e:gpt2a} in Section~\ref{sec:gflow},
and then consider the observables in Section~\ref{sec:focc}.
The analysis is based on a formula for $P$ obtained in
Section~\ref{sec:Pformula}.

\subsubsection{Preliminary identities}
\label{sec:Pformula}

Since $e^{\Lcal_C}$ reduces the dimension of a monomial in the fields,
$e^{\Lcal_C} : \Vcal \to \Vcal$, and since $\LT_X$ acts as the identity
on $\Vcal$, it follows that
\begin{equation}
\label{e:LTELT}
    \LT_X e^{\Lcal_C} \LT_X = e^{\Lcal_C} \LT_X.
\end{equation}
The following lemma gives the formula we use to compute $P$.

\begin{lemma}
\label{lem:Palt}
For $x\in \Lambda$, for any local polynomial $V$, and for covariances $C,w$,
\begin{align}
    P_x
\label{e:Palt0}
    &=
    \frac 12
    \sum_{y \in \Lambda}
    \big(
    \LT_{x}
    F_{\pi,w+C} (e^{\Lcal_C} V_x,e^{\Lcal_C} V_y) -
    e^{\Lcal_C} \LT_{x}  F_{\pi,w} (V_x,V_y)
    \big)
    .
\end{align}
\end{lemma}

\begin{proof}
The definition of $P$ is given in \eqref{e:PdefF}, namely
\begin{equation}
\label{e:PdefFbis}
    P_x = \LT_x e^{\Lcal_C} W_j(V,x)
    + \frac 12 \LT_x
    F_{\pi,C}
    (e^{\Lcal_C} V_x,e^{\Lcal_C} V(\Lambda)).
\end{equation}
By the definition of $W_j$ in \eqref{e:WLTF}, this can be rewritten as
\begin{equation}
\label{e:Pxformula}
    P_x = \frac 12 \LT_x \big(
    e^{\Lcal_C}
    (1-\LT_x)F_{\pi,w}( V_x, V(\Lambda))
    + F_{\pi,C} (e^{\Lcal_C} V_x,e^{\Lcal_C} V(\Lambda))
    \big).
\end{equation}
Application of \eqref{e:LTELT} in \eqref{e:Pxformula} gives
\begin{align}
    P_x
    & =
    \frac 12 \LT_x \big(
    e^{\Lcal_C}
    F_{\pi,w}( V_x, V(\Lambda))
    + F_{\pi,C} (e^{\Lcal_C} V_x,e^{\Lcal_C} V(\Lambda))
    \big)
    \nnb
    &\qquad
    -
    \frac{1}{2}
    e^{\Lcal_C}
    \LT_x F_{\pi,w}( V_x, V(\Lambda))
    .
\label{e:Pxformula1}
\end{align}

By the definition of $F$ in \eqref{e:FCAB}, for polynomials $A,B$ in
the fields,
\begin{align}
    F_{w+C}(e^{\Lcal_C}A,e^{\Lcal_C}B)
    &=
    e^{\Lcal_C}
    e^{\Lcal_w}
    \big(e^{-\Lcal_{w}}A\big)
    \big(e^{-\Lcal_{w}} B\big)
    -(e^{\Lcal_C}A)(e^{\Lcal_C}B)
    \nnb
    &=
    e^{\Lcal_C}
    F_{w}( A, B) +
    e^{\Lcal_C} \left(AB \right)
    - (e^{\Lcal_C}A)(e^{\Lcal_C}B)
    \nnb
    &=
    e^{\Lcal_C}
    F_{w}( A, B) + F_{C}(e^{\Lcal_C}A,e^{\Lcal_C}B)
    .
\label{e:EFAB}
\end{align}
By \eqref{e:Fpi},
\eqref{e:EFAB} extends to
\begin{equation}
\label{e:EthF}
    e^{\Lcal_C} F_{\pi,w} (A,B)
    +
    F_{\pi,C}(e^{\Lcal_C} A, e^{\Lcal_C} B)
    =
    F_{\pi,w+C}(e^{\Lcal_C} A, e^{\Lcal_C} B)
    .
\end{equation}
With \eqref{e:EthF}, \refeq{Pxformula1} gives
\begin{align}
    P_x
    &=
    \frac 12
    \LT_{x} F_{\pi,w+C} (e^{\Lcal_C} V_x,e^{\Lcal_C} V(\Lambda))
    -
    \frac 12
    e^{\Lcal_C}\LT_{x} F_{\pi,w} ( V_x, V(\Lambda))
    ,
\end{align}
and the right-hand side is equal to the right-hand side of \eqref{e:Palt0}.
\end{proof}

The first step in the evaluation of the right-hand side of \eqref{e:Palt0}
is to compute $F_w(V_x,V_y)$.
We do this with the following lemma.
Given a symmetric covariance $w$, and
polynomials $V',V''$, we define
\begin{equation}
\label{e:Lcallrdef}
    V'\Lcallr_wV''
    =
    \sum_{u,v \in \Lambda} w_{uv}
    \left(
    \frac{\partial V'}{\partial \phi_u} \frac{\partial V''}{\partial \bar\phi_v}
    +
    \frac{\partial V'}{\partial \phi_v} \frac{\partial V''}{\partial \bar\phi_u}
    +
    \frac{\partial V'}{\partial \psi_u} \frac{\partial V''}{\partial \bar\psi_v}
    +
    \frac{\partial V'}{\partial \psi_v} \frac{\partial V''}{\partial \bar\psi_u}
    \right).
\end{equation}
For $n \ge 2$, we define $V' (\Lcallr_w)^n V''$
analogously as a sum over $u_1,v_1,\ldots,u_n,v_n$, with
$n$ derivatives acting on each of $V'$ and $V''$, with
$n$ factors $w$
as in \eqref{e:Lcallrdef}.

\begin{lemma}
\label{lem:Fexpand}
For $x,y\in \Lambda$, for a local polynomial $V$ of
degree $A$,
and for a covariance $w$,
\begin{equation}
\label{e:Fexpand1Vpt}
    F_{w}(V_x,V_y) = \sum_{n=1}^A  \frac{1}{n!} V_x (\Lcallr_{w})^n V_y.
\end{equation}
\end{lemma}

\begin{proof}
By \eqref{e:FCAB},
\begin{align}
    F_{w}(V_x,V_y) &
    = e^{\Lcal_{w}}
    \big(e^{-\Lcal_{w}}V_x\big)
    \big(e^{-\Lcal_{w}}V_y \big) - V_x V_y
\label{e:Fpiw}
    .
\end{align}
The Laplacian can be written as a sum of three contributions,
one acting only on $V_x$, one only on $V_y$,
and the cross term \eqref{e:Lcallrdef}.  The first two terms are
cancelled by the operators $e^{-\Lcal_w}$ appearing in \eqref{e:Fpiw},
leading to
\begin{equation}
    F_{w}(V_x,V_y)
    = V_x e^{\Lcallr_{w}} V_y - V_x V_y.
\end{equation}
Expansion of the exponential then gives \eqref{e:Fexpand1Vpt}.
\end{proof}

\subsubsection{Localisation operator}

The computation of the flow equations requires the calculation of $P$,
which involves the operator $\LT$ as indicated in
Lemma~\ref{lem:Palt}.  An extensive discussion of the operator $\LT$
is given in \cite{BS-rg-loc}, and
\cite[Example~\ref{loc-ex:LTmunu}]{BS-rg-loc} gives some sample
calculations involving $\LT$.  Given the specifications listed in
Section~\ref{sec:loc-specs}, it follows from the definition of $\LT$
that
\begin{equation} \label{e:LTtau2}
  \LT_{x}\left[ \tau_{a_1 b_1}\tau_{a_2 b_2} \right] = \tau_{x}^2
  ,
\end{equation}
and we use this repeatedly in our calculation of $\gpt$ below.  Also,
for the calculation of $\lambdapt$ and $\qpt$, we use the fact that
the monomials $\sigmaa \Delta \bar{\phi}$ and $\sigmab \Delta \phi$
are annihilated by $\LT$.

We do not provide the details of the calculation of $\nupt$, $\ypt$,
and $\zpt$ here.  As mentioned previously, their flow in
\eqref{e:nupta}--\eqref{e:zpta} has been computed using a Python
computer program.  To help explain the nature of the terms that arise
in these equations, we note the following facts about $\LT$, which
extend \cite[Example~\ref{loc-ex:LTmunu}]{BS-rg-loc} and which are
employed by the Python program.  First, monomials of degree higher
than $4$ are annihilated by $\LT$.  Less trivially, suppose that $q:
\Lambda \to \C$ has range strictly less than the period of the torus
and that it satisfies, for some $q^{(**)} \in \C$,
\begin{equation}
  \label{e:qprop1}
  \sum_{x \in \Lambda} q (x) x_{i} = 0,
  \quad\quad
  \sum_{x \in \Lambda} q (x) x_{i}x_{j}
  = q^{(**)} \delta_{i,j},
  \quad\quad\quad
  i,j \in \{1,2,\dotsc ,d \}.
\end{equation}
Then
\begin{align}
  \label{e:LTF3}
  \LT_{x}
  \left[
    \sum_{y \in \Lambda} q (x-y) \tau_{y}
  \right]
  &=
  q^{(1)}\tau_{x}
  +
  q^{(**)} (\tau_{\nabla\nabla,x}-\tau_{\Delta,x})
  ,
  \\
  \label{e:LTF4}
  \LT_{x}
  \left[
    \sum_{y \in \Lambda} q (x-y)
    (
    \tau_{xy} + \tau_{yx}
    )
  \right]
  &=
  2 q^{(1)}\tau_{x}
  +
  q^{(**)} \tau_{\Delta,x}
  .
\end{align}
In particular, the coefficients $\theta,\sigma,\zeta$ of
\refeq{betadef} and \refeq{sigzetadef} have their origin in
\eqref{e:LTF3}--\eqref{e:LTF4}.
To simplify the result of the computation,
we have also used the elementary properties that for any
$q: \Lambda \to \C$,
\begin{equation}
  \sum_{x\in\Lambda} (\nabla^e q)_x = 0,
  \quad \sum_{x\in\Lambda} (\Delta q)_x = 0,
\end{equation}
as well as the fact that $\Delta x_1^2 = -2$, which,  by summation by parts, implies that
\begin{equation}
   \sum_{x\in\Lambda} (\Delta q)_x x_1^2 = - 2 \sum_{x\in\Lambda} q_x = -2 q^{(1)}.
 \end{equation}

\subsubsection{Flow of \texorpdfstring{$g$}{g}}
\label{sec:gflow}

We now prove the flow equation \eqref{e:gpt2a} for $g_{\pt}$.
As in Lemma~\ref{lem:Fexpand}, we write
\begin{equation}
\label{e:Fexpand1Vptbis}
    F_{w}(V_x,V_y) = F_{xy} = \sum_{n=1}^4 F_{n;xy}
\quad \text{with}
\quad
    F_{n;xy} = \frac{1}{n!} V_x (\Lcallr_{w})^n V_y.
\end{equation}
The main work lies in
proving the following lemma.

\begin{lemma}
\label{lem:Fexplicit}
The $\tau_x^2$ term in
$\sum_{y \in \Lambda} \LT_{x} F_{xy}$
is equal to
\begin{align}
\label{e:Fformula}
    \big(
    16g^{2} w^{(2)} + 8g\nu w^{(1)}
    \big)\tau_{x}^{2}
    .
\end{align}
\end{lemma}

Before proving Lemma~\ref{lem:Fexplicit}, we first note that it
implies \eqref{e:gpt2a}.

\begin{proof}[Proof of \eqref{e:gpt2a}.]  By
Lemmas~\ref{lem:Fexplicit} and \ref{lem:EV},
the $\tau_x^2$ term in $e^{\Lcal_C} \LT_{x} F_{w} (V_{x}, V_y )$ is given by
\begin{align}
    \label{e:ExLTF}
    \big(
    16g^{2} w_{x,y}^2 + 8g\nu w_{x,y}
    \big)\tau_{x}^{2}
    .
\end{align}
Also, by \eqref{e:EV}, $e^{\Lcal_C} V$ is equal to $V$
with the coefficient $\nu$ replaced by
\begin{align}
    \label{e:nuplus}
    &
    \nu_{+}
    =
    \nu + 2g C_{0,0}
,
\end{align}
so by Lemma~\ref{lem:Fexplicit} the $\tau^2$ term
in $\sum_{y \in \Lambda}\LT_{x} F_{w+C} (e^{\Lcal_C} V_{x},e^{\Lcal_C} V_y)$ is given by
\begin{align}
    \big(
    16g^{2} w_+^{(2)} + 8g\nu_{+} w_{+}^{(1)}
    \big)\tau_{x}^{2}
    ,
    \label{e:LTFplus}
\end{align}
where $w_+=w+C$.
By Lemma~\ref{lem:Palt}, the $\tau_x^2$ term in $P_x$ is
therefore equal to
\begin{align}
    \label{e:P2}
    \big(
    8g^{2} \delta[w^{(2)}] + 4g \delta[\nu w^{(1)}]
    \big)\tau^{2}_x
    .
\end{align}
With the formula $\Vpt=e^{\Lcal_C} V -P$
from \eqref{e:Vptdef}, this implies that
\begin{align}
    \gpt
    &
    =
    g
    -
    8 g^{2} \delta[w^{(2)}]
    - 4 g \delta[\nu w^{(1)}],
    \label{e:gpt2b}
\end{align}
which is \eqref{e:gpt2a}.
\end{proof}

\begin{proof}[Proof of Lemma~\ref{lem:Fexplicit}.]
We compute the $\tau_x^2$ term in $\LT_{x}
F_{n;x,y}$ for $n=1,2,3,4$.  Since $F_{4;x,y}$ has degree zero
it contains no $\tau_x^2$ term,
so it suffices to consider $n=1,2,3$.
The observables play no role in this discussion, and we can let
\begin{equation}
\label{e:V}
    V = g \tau^{2} + \nu\tau + z\tau_{\Delta} + y\tau_{\nabla\nabla}.
\end{equation}
To compute $F_{n;x,y}$ for $n=1,2,3$, we take
the terms in \eqref{e:V} into account sequentially, starting with
$g\tau^2$, then $\nu \tau$, then $z\tau_\Delta$, and finally $y\tau_{\nabla\nabla}$.

\medskip \noindent
\emph{$\tau^2$ term.}
We first study
\begin{equation}
    F_{n;xy}
    =
    \frac{1}{n!}
    A_x (\Lcallr_{w})^n A_{y}
    \quad \text{with} \quad
\label{e:AFeynman-def}
    A = g\tau^{2}.
\end{equation}
By \eqref{e:AFeynman-def}, $F_{1}$ is a polynomial whose terms are
degree $6$ and therefore $\LT_{x} F_{1;xy} = 0$.  Also, $F_{3;xy}$
is a polynomial whose monomials have degree $2$, and therefore we need
not calculate them here.  Thus we need only compute the $\tau^2$
contribution to $F_{2;xy}$.

To make contact with Section~\ref{sec:closed-loops}, we replace
$A_{x}$ and $A_{y}$ by
\begin{equation}
    A_{12} = g\tau_{a_{1}a_{1}} \tau_{a_{2}a_{2}},
    \quad
    A_{34} = g\tau_{a_{3}a_{3}} \tau_{a_{4}a_{4}}.
\end{equation}
The labels  $1,2,3,4$ help enumerate terms
that result from carrying out the contractions in $F$, but after the
enumeration many of these terms become the same when we return to the
case at hand by setting
\begin{equation}\label{e:coincident}
    a_{1}=a_{2}=x,
    \quad
    a_{3}=a_{4}=y.
\end{equation}
We represent $A$ by
\begin{center}
\includegraphics[scale = 0.6]{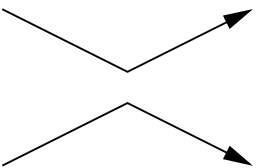}
\quad .
\end{center}

The diagrams for $F_{2;xy}$ are given by
\begin{center}
\includegraphics[scale = 0.6]{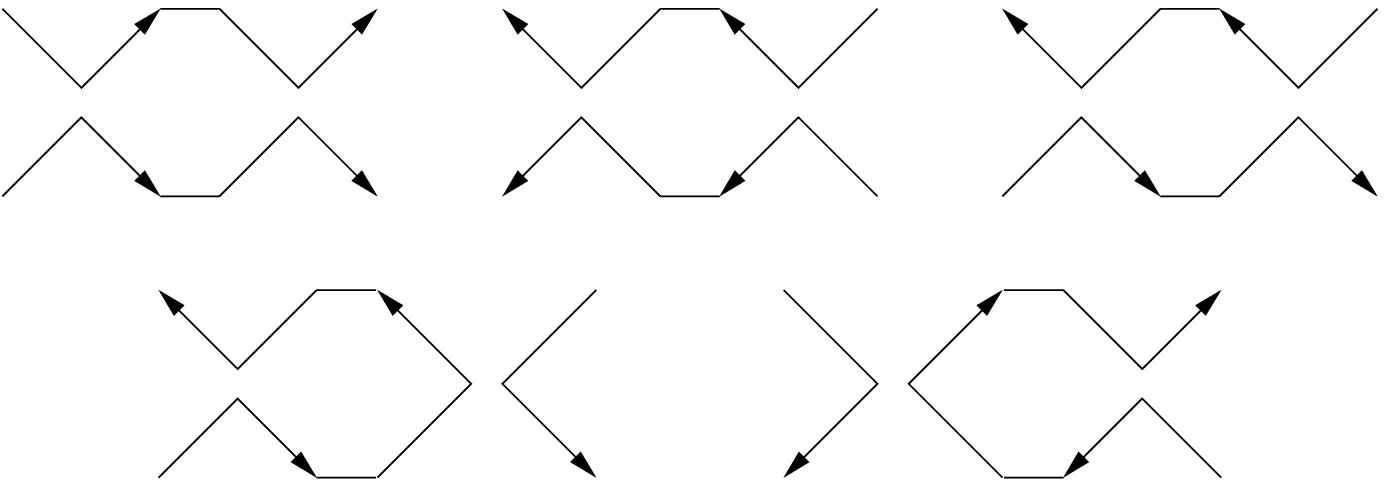},
\end{center}
as well as the diagram
\begin{center}
\includegraphics[scale = 0.6]{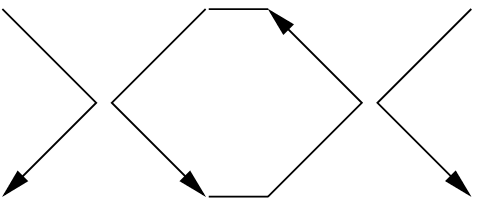}
\end{center}
which contains a closed loop.  The latter vanishes,
because it arises, for example, from
$\tau_{1} (\Lcal^2 \tau_2\tau_3) \tau_4$ which is
$0$ by Lemma~\ref{lem:noloops}.
We claim that the five diagrams without closed loops amount to
\begin{equation}
    \label{e:F2}
    F_{2;xy}
    =
    g^{2}
    w^{2} (x,y)
    \big(
    2\tau_{xy}^{2} + 2\tau_{yx}^{2} + 4\tau_{xy}\tau_{yx} +
    4\tau_{x}\tau_{y}  + 4\tau_{x}\tau_{y}
    \big).
\end{equation}
As a preliminary observation note that the prefactor of $\frac{1}{2!}$
in \eqref{e:AFeynman-def} cancels the $2!$
identical terms that arise from the order of the two contractions
in applying \eqref{e:contraction} twice,
so for each diagram we only count matchings: how many
ways out-legs can be matched to in-legs.
The five terms correspond to
the five diagrams.  These arise as follows:
\\
First and second diagrams: each diagram has two matchings.
\\
Third diagram: four  matchings.
\\
Fourth and fifth diagrams: each diagram has four  matchings.
\\
Since all terms on the right-hand side of \eqref{e:F2} are fourth
order in the fields, it is immediate
from \eqref{e:LTtau2}
that the $\tau^2$ contribution to
$\LT_{x} F_{2;xy}$ is given by
\begin{equation}
\label{e:LTF2}
    \LT_{x} F_{2;xy} = 16g^2 w^2_{x,y} \tau_x^2
    .
\end{equation}

\medskip \noindent
\emph{$\tau$ term.}
Now we consider the additional terms that arise when we add a $\nu
\tau$ term so that
\begin{equation}
    \label{e:AxAynu}
    A = g\tau^{2} + \nu \tau.
\end{equation}
The additional terms in $F_{2,xy}$ are not needed since they are of
degree $2$, and there are no additional terms in $F_{3,xy}$.
Repeating the calculations for $F_{1,xy}$ with the extra term in $A$
we obtain the additional diagrams
\begin{center}
\includegraphics[scale = 0.6]{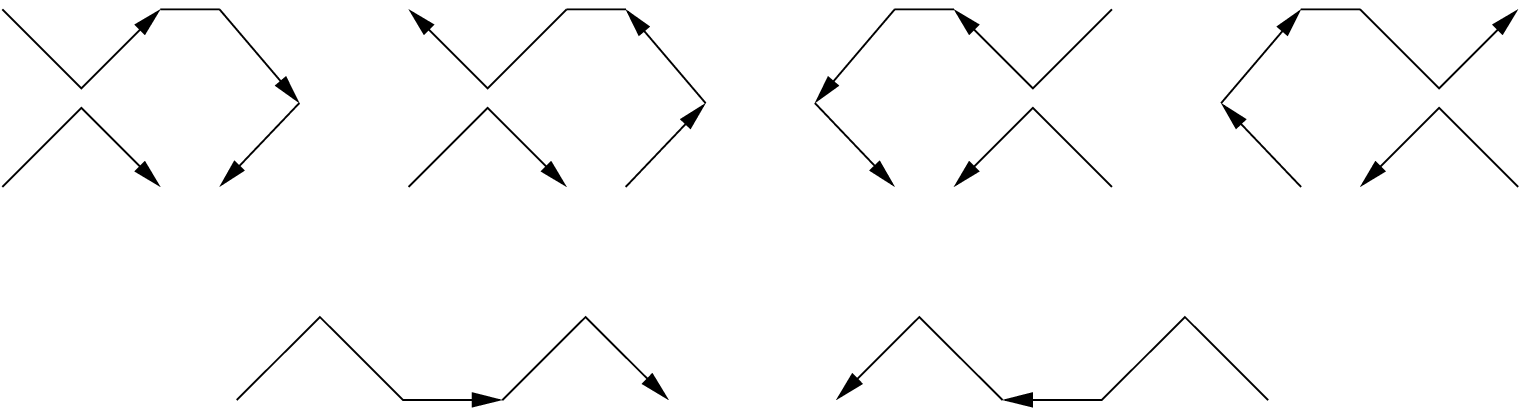}.
\end{center}
Therefore $F_{1,xy}$ has the additional terms
\begin{align}
    \label{e:F1-mass}
    &
    2g\nu
    \tau_{x}
    w_{x,y}
    (\tau_{xy} + \tau_{yx})
    +
    2g\nu
    \tau_{y}
    w_{x,y}
    (\tau_{yx} + \tau_{xy})
    +
    \nu^{2}
    w_{x,y}
    (\tau_{xy} + \tau_{yx}).
\end{align}
Thus, by \eqref{e:LTtau2},
the additional $\tau^2$ contribution that arises here after
localisation  is:
\begin{align}
    \label{e:loc-g-nu}
    8g\nu
    w_{x,y}
    \tau_{x}^{2}
    .
\end{align}

\medskip \noindent
\emph{$\tau_{\Delta}$ term.}
Now we consider the additional degree four terms that arise by adding
$z\tau_{\Delta}$ to $A$, with $\tau_\Delta$ defined in \eqref{e:addDelta}.
These degree four terms arise from contractions between $\tau_x^2$
and $\tau_{\Delta,y}$, and between $\tau_{\Delta,x}$ and $\tau_y^2$.
After localisation at $x$, these yield contributions involving
$(\Delta_y w_{x,y})\tau_x^2$ and $(\Delta_x w_{x,y})\tau_x^2$.
These both vanish after summation over $y \in \Lambda$.
Thus there is no
contribution to $ \sum_{y} \LT_{x}F_{1,xy}$ arising from the
$\tau_\Delta$ term.

\medskip \noindent
\emph{$\tau_{\nabla\nabla}$ term.}
Now we consider the additional degree four terms that arise by adding
$y\tau_{\nabla\nabla}$ to $A$, with $\tau_{\nabla\nabla}$ defined in
\eqref{e:taunabnabdef}.
These contributions are similar to those for $\tau_\Delta$, and
after localisation at $x$, produce contributions involving
$\sum_{e \in \units}(\nabla_x^e \nabla_y^e w_{x,y})\tau_x^2$,
which vanishes after summation over $y \in \Lambda$.
Thus there is no
contribution to $\sum_{y} \LT_{x} F_{1,xy}$ arising from the
$\tau_{\nabla\nabla}$ term.

\medskip \noindent
The proof of Lemma~\ref{lem:Fexplicit} is now completed
by combining \eqref{e:LTF2} and \eqref{e:loc-g-nu}.
\end{proof}

\subsubsection{Flow of \texorpdfstring{$\lambda,q$}{lambda,q}}
\label{sec:focc}

We now prove the following lemma, which implies the flow equations
\eqref{e:lambdapt2}--\eqref{e:qpt2}.

\begin{lemma}
\label{lem:Pobs}
For $j + 1 < j_{\pp\qq}$,
the observable part of $P = P_{j}$ as defined in \eqref{e:PdefF} is
given by
\begin{equation}
\label{e:Pbbdobs-app}
    \pi_* P_x =
    - \delta[\nu w^{(1)}]
    (\lambdaa \sigma \bar\phi_a \1_{x=a}+ \lambdab \bar\sigma  \phi_b \1_{x=b})
    + \frac 12 C_{ab} \lambdaa \lambdab \sigma \bar\sigma  (\1_{x=a}+ \1_{x=b})
    ,
\end{equation}
while for $j + 1 \ge j_{\pp\qq}$
\refeq{Pbbdobs-app} holds with the first term on the right-hand side
replaced by zero.
\end{lemma}

\begin{proof}
The distinction involving $j_{\pp\qq}$ arises due to the change in
$d_+$ discussed in Section~\ref{sec:loc-specs}, which stops
$\lambdaa,\lambdab$ from evolving above the coalescence scale.
Throughout the proof, we consider only the more difficult case of $j +
1 < j_{\pp\qq}$.

We consider the effect on $P$ of adding the observable terms $\pi_*V$
into $V$.  The Laplacian annihilates the $\sigmaa\sigmab$ term and it
cancels in the subtraction in \eqref{e:Fpiw}, so can be dropped
henceforth from $V$.  Thus we wish to compute the new contributions
that arise after adding the observable terms
\begin{equation}
    A'_{x}
    =
    - \lambdaa \,\sigmaa \bar{\phi}_{x}\1_{x = \pp}
    -
    \lambdab \,\sigmab \phi_{x}\1_{x = \qq}
\end{equation}
to $A_{x}$ and $A_y$.  We will see, in particular, that there is no
contribution from the observables to the flow of non-observable
monomials.

Recall that $d_+=[\phi]=1$ in the definition of $\LT$ restricted to
$\pi_*\Ncal$.  We first consider the $\pi_a\Ncal$ and $\pi_b \Ncal$
terms.  Writing $F = \sum_{n=1}^4 \frac{1}{n!}  F_{n}$ as before, we
need only consider the $n=1$ term because the observables are degree
one polynomials in $(\phi,\bar\phi )$.  Contractions with $g\tau^{2}$
give rise to monomials that are annihilated by $\LT$ and therefore
make no contribution.  Contractions with $\nu \tau$ produce
\begin{center}
\includegraphics[scale = 0.6]{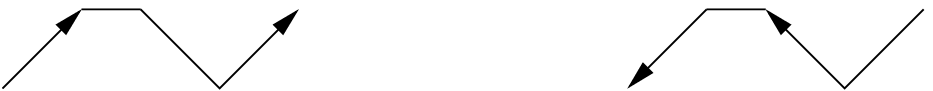},
\end{center}
and, according to \eqref{e:Fpi}, the contribution of these diagrams to
$F_{\pi,w}(A_x, A_y)$ is
\begin{equation}
\label{e:lambdaFterm}
    A'_{x} \Lcallr_{w} (\pi_{\varnothing}\nu \tau_{y}) +
    (\pi_{*}A'_{x}) \Lcallr_{w} (\nu \tau_{y})
=
    - 2 \nu \, w_{x,y}
    \big(
    \lambdaa\sigmaa \bar{\phi}_{y}\1_{x=\pp}
    +
    \lambdab\sigmab \phi_{y}\1_{x=\qq}
    \big).
\end{equation}
These same diagrams also classify contractions between the observables
and $z \tau_{\Delta}$ or $y\tau_{\nabla\nabla}$, but in this case make
no contributions to $\LT_{x}\sum_{y\in \Lambda} F_{xy}$ since, e.g.,
$\sigmaa \Delta \bar{\phi}$ is annihilated by $\LT$.  Thus
\eqref{e:lambdaFterm} constitutes the new terms arising from
contractions between observable and non-observable terms in
$F_{\pi,w}(A_x,A_y)$.

Next, we consider the $\pi_{ab} \Ncal$ term.
The contraction of the $\lambda$ terms in
$A_x$ with those in $A_y$ results in
\begin{center}
\includegraphics[scale = 0.6]{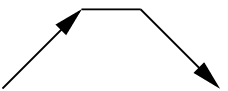}
\end{center}
which contributes
\begin{equation}
    \lambdaa\lambdab \,\sigmaa \sigmab
    \big(
    \1_{x=\pp}\1_{y=\qq}w_{x,y}
    +
    \1_{x=\qq}\1_{y=\pp} w_{y,x}
    \big).
\end{equation}
Using $w_{a,b}=w_{b,a}$, and using \eqref{e:Fexpand1Vptbis},
this makes a contribution
\begin{equation}
\label{e:piabF}
    \lambdaa\lambdab \,\sigmaa \sigmab w_{\pp,\qq} \,
    (\1_{x =\pp} + \1_{x = \qq})
\end{equation}
to
$\LT_{x}\sum_{y\in\Lambda}F_{1;x,y}(A_{x},A_{y})$.
By Lemma~\ref{lem:Palt}, we find that the contribution to $P_{x}$ is
\begin{align}
    &
    - \delta[\nu w^{(1)}] \,
    \big(
    \lambdaa \sigmaa \bar{\phi}_{x} \1_{x = \pp}
    +
    \lambdab \sigmab \phi_{x} \1_{x = \qq}
    \big)
    +
    \half
    \lambdaa\lambdab \,\sigmaa \sigmab \,\delta[w_{\pp,\qq}]\,
    (\1_{x =\pp} + \1_{x = \qq})
    .
\end{align}
Since $\delta[w_{\pp,\qq}] = C_{j+1;\pp,\qq}$,
this completes the proof.
\end{proof}

\section{Analysis of flow equations}
\label{sec:Greekpfs}

In this section, we prove
Propositions~\ref{prop:transformation}--\ref{prop:rg-pt-flow}.
This requires details of the specific covariance decompositions
we use.  In
Section~\ref{sec:Cdecomp}, we define the covariance decompositions,
list their important properties, and use those properties
to obtain estimates
on the coefficients \refeq{betadef}--\refeq{sigzetadef} of the
flow equations.  Then we prove
Propositions~\ref{prop:transformation}--\ref{prop:rg-pt-flow}
in Sections~\ref{sec:pf-transformation}--\ref{sec:pf-alpha},
respectively.

\subsection{Decomposition of covariance}
\label{sec:Cdecomp}

\subsubsection{Definition of decomposition}

Let $d>2$.
We begin by describing the specific finite-range
decomposition of the covariance
$[-\Delta_{\Zd}+m^2]^{-1}$ we use,
from \cite{Baue13a} (see also \cite{Bryd09,BGM04}).
Recall from
\cite[Example~1.1]{Baue13a} that for each $m^2 \ge 0$ there is a function
$\phi_t^*(x,y;m^2)$
defined for $x,y \in \Z^d$ and $t>0$
such that
\begin{equation} \label{e:decomptZd}
  [-\Delta_{\Z^d} +m^2]^{-1}_{x,y}
  = \int_0^\infty \phi_t^*(x,y;m^2) \, \frac{dt}{t}.
\end{equation}
The function $\phi_t^*$ is positive definite as a function of $x,y$,
has the finite-range property  that $\phi_t^*(x,y;m^2)=0$ if $|x-y|>t$
(this specific range can be achieved by rescaling in $t$),
and is Euclidean invariant (this can be seen, e.g., from \cite[(3.19)]{Baue13a}).
To obtain $\phi_t^*$ as a well-behaved function of $m^2$, it is necessary
to restrict to a finite interval $m^2\in [0,\bar m^2]$ and we make this restriction
in the following.
Further properties of $\phi_t^*$ are recalled in the proof
of Proposition~\ref{prop:Cdecomp} below.
Let
\begin{equation} \label{e:Cdef}
  C_{j;x,y}= \left\{\begin{aligned}
      &\int_{0}^{\frac{1}{2} L}  \phi_{t}^*(x,y; m^2) \; \frac{dt}{t}
      &\quad& (j=1)\\
      &\int_{\frac{1}{2} L^{j-1}}^{\frac{1}{2} L^{j}}  \phi_{t}^*(x,y; m^2) \; \frac{dt}{t}
      && (j\ge 2).
    \end{aligned}\right.
\end{equation}
Each $C_j$ is a positive-definite $\Zd \times \Zd$ matrix,
is Euclidean invariant,
has the finite-range property
\begin{equation}
\lbeq{Cjfinran}
    C_{j;x,y} = 0 \quad \text{if $|x-y| \geq \frac{1}{2} L^j$},
\end{equation}
and, by construction,
\begin{equation}
    C= [-\Delta_\Zd +m^2]^{-1} = \sum_{j=1}^\infty C_j.
\end{equation}
This is the covariance decomposition we employ in \refeq{ZdCj}.

Next, we adapt \refeq{Cdef} to obtain a decomposition for
the torus $\Lambda = \Z^d / L^N \Z^d$.
Let $L,N>0$ be integers and $m^2 \in (0,\bar m^2)$.
By \refeq{Cjfinran}, $C_{j;x,y+L^Nz} = 0$ for $j<N$, $|x-y| < L^N$,
and nonzero $z \in \Z^d$,
and thus
\begin{equation}
     C_{j;x,y} = \sum_{z \in \Z^d} C_{j;x,y+zL^N} \quad \text{for $j<N$}.
\end{equation}
We therefore can and do regard $C_{j}$ either as a $\Zd\times\Zd$ matrix
or as a $\Lambda \times \Lambda$ matrix if $j<N$.
We also define
\begin{equation} \label{e:CNNdef}
     C_{N,N;x,y} = \sum_{z \in \Z^d} \sum_{j=N}^\infty C_{j;x,y+zL^N}.
\end{equation}
Then $C_j$ and $C_{N,N}$ are Euclidean invariant on $\Lambda$
(i.e., invariant under automorphisms $E: \Lambda \to\Lambda$
as defined in Section~\ref{sec:supersymmetry}).
Since
\begin{equation}
  [-\Delta_\Lambda + m^2]^{-1}_{x,y} = \sum_{z \in \Z^d} [-\Delta_{\Z^d}+m^2]^{-1}_{x,y+zL^N},
\end{equation}
it also follows that
\begin{equation}
    [-\Delta_\Lambda + m^2]^{-1}
    = \sum_{j=1}^{N-1} C_{j} + C_{N,N}.
\end{equation}
Therefore the effect of the torus is concentrated in the term $C_{N,N}$.
This is the decomposition used in \refeq{NCj}.

\subsubsection{Properties of decomposition}

The
following proposition provides estimates on the finite-range decomposition
defined above.
In its statement, given a multi-index $\alpha=(\alpha_1,\dotsc,\alpha_d)$,
we write
$\nabla_x^\alpha=\nabla_{x_1}^{\alpha_1} \dotsb \nabla_{x_d}^{\alpha_d}$
 where
$\nabla_{x_k}$ denotes the
finite-difference operator defined by $\nabla_{x_k}f(x,y)=f(x+e_k,y)-f(x,y)$.
The number $[\phi]$ is equal to $\frac 12
(d-2)$ as in \refeq{phidim}.

\begin{prop}
\label{prop:Cdecomp}
  Let $d > 2$, $L\geq 2$, $j \ge 1$, $\bar m^2 >0$.
  \begin{enumerate}[(a)]
  \item
  For multi-indices $\alpha,\beta$ with
  $\ell^1$ norms $|\alpha|_1,|\beta|_1$ at most
  some fixed value $p$,
  and for any $k$, and for $m^2 \in [0,\bar m^2]$,
  \begin{equation}
    \label{e:scaling-estimate}
    |\nabla_x^\alpha \nabla_y^\beta C_{j;x,y}|
    \leq c(1+m^2L^{2(j-1)})^{-k}
    L^{-(j-1)(2[\phi]+(|\alpha|_1+|\beta|_1))},
  \end{equation}
  where $c=c(p,k,\bar m^2)$ is independent of $m^2,j,L$.
  The same bound holds for $C_{N,N}$ if
  $m^2L^{2(N-1)} \ge \varepsilon$ for some $\varepsilon >0$,
  with $c$ depending on $\varepsilon$ but independent of $N$.

  \item
    For $j>1$, the covariance $C_j$ is differentiable in $m^2 \in (0,\bar m^2)$, right-continuous
    at $m^2=0$, and there is a constant $c>0$ independent of $m^2,j,L$ such that
    \begin{equation}
      \label{e:decomp-deriv}
      \left|\frac{\partial}{\partial m^2} C_{j;x,y}\right|
      \leq c(1+m^2L^{2(j-1)})^{-k}
      \begin{cases}
        L^{j} &(d=3)\\
        \log L & (d=4)\\
        L^{-(d-4)(j-1)} & (d > 4).
      \end{cases}
    \end{equation}
    Furthermore,
    $C_1$ is continuous in $m^2 \in (0,\bar m^2)$ and right-continuous at $m^2 = 0$, and
    $C_{N,N}$ is continuous in the open interval $m^2 \in (0,\bar m^2)$. 

  \item
    Let $m^2=0$.
    There exists a smooth function
    $\rho: [0,\infty) \to [0,\infty)$
    with
    $\int_0^\infty \rho(t) \, dt = 1$, such that the function $c_0: \R^d \to \R$
    defined by its Fourier transform
    $\hat c_0(\xi) = |\xi|^{-2} \int_{L^{-1}|\xi|}^{|\xi|} \rho(t) \, dt$
    is smooth with compact support, and, as $j\to\infty$,
    \begin{equation}
      C_{j;x,y} = c_j(x-y) + O(L^{-(d-1)(j-1)})
      \quad \text{for $m^2 =0$},
    \end{equation}
    where $c_j(x) = L^{-(d-2)j}c_0(L^{-j}x)$.
  \end{enumerate}
\end{prop}

\begin{proof}
  We use the results of \cite[Example 1.1]{Baue13a}.\\
(a,b)
  Let $p,k \in \N$. For any $\alpha, \beta$ with $|\alpha|_1,|\beta|_1 \leq p$,
  \cite[(1.35)]{Baue13a} implies that there is $c = c(p,k)$ such that
  \begin{align}
    \label{e:decomp-phi-est}
    |\nabla_x^\alpha\nabla_y^\beta \phi_t^*(x,y;m^2)|
    &\leq c(1+m^2 t^{2})^{-k} t^{-(2[\phi]+|\alpha|_1+|\beta|_1)}, \\
    \label{e:decomp-phi-deriv-est}
    \left|\ddp{}{m^2}
    \phi_t^*(x,y;m^2) \right|
    &
    \leq c(1+m^2t^{2})^{-k} t^{-(2[\phi]
    -2)
    }
  \end{align}
  (using $\half d > 1$ in \cite[(1.36)]{Baue13a} for the second bound).
  Consider first the case $j>1$.  In this case, we restrict
  to $t\in [\frac 12 L^{j-1},\frac 12 L^j]$ and obtain upper bounds
  in \refeq{decomp-phi-est}--\refeq{decomp-phi-deriv-est} by replacing $t^2$
  by $L^{2(j-1)}$.  Substitution of the resulting estimates into
  \refeq{Cdef} imply \eqref{e:scaling-estimate}--\eqref{e:decomp-deriv}
  for $j > 1$,
  with constants independent of $L$.
  For example, $\log L$ in \refeq{decomp-deriv} arises as
  \begin{equation}
    \int_{\frac 12 L^{j-1}}^{\frac 12 L^j} \frac{dt}{t} = \log L
    .
  \end{equation}
  Moreover, since $\phi_t^*(x,y;m^2)$ is continuous in $m^2$ at $m^2=0^+$
  and bounded for
  $t\in [\frac 12 L^{j-1},\frac 12 L^j]$ by \refeq{decomp-phi-est},
  for $j>1$
  the claimed right-continuity of $C_{j;x,y}$
  at $m^2=0^+$ is a straightforward consequence
  of the dominated convergence theorem.

  For $j=1$, the bound \eqref{e:decomp-phi-est} needs to be improved. To this end,
  we use the discrete heat kernel $p_t(x,y) = (\delta_y, e^{\Delta t} \delta_x)$. Since $e^{\Delta t}$
  is a contraction on $L^2(\Z^d)$ and $\delta_x \in L^2(\Z^d)$, it follows that $p_t(x,y)$ is uniformly bounded,
  i.e., $p_t(x,x) \leq ct^{-\alpha/2}$ with $\alpha=0$.
  Thus \cite[Theorem 1.1]{Baue13a} and \refeq{decomp-phi-est} imply that
  \begin{equation}
    \label{e:decomp-phi-est-j1}
    |\phi_t^*(x,y;m^2)|
    \leq c(t^{-2[\phi]} \wedge t^2)
    .
  \end{equation}
  It follows that
  \begin{equation}
    |C_{1;x,y}|
    = \left| \int_0^{\frac 12L} \phi_t^*(x,y;m^2) \; \frac{dt}{t}\right|
    \leq c \int_0^1 t^2 \; \frac{dt}{t} + c \int_1^{\frac 12 L} t^{-2[\phi]} \; \frac{dt}{t} \leq \mathrm{const}.
  \end{equation}
  This proves \refeq{scaling-estimate} for $j=1$ with $\alpha=\beta=0$, and
  the estimates for $|\alpha|_1, |\beta|_1 \leq p$ are an immediate consequence because the discrete difference operator is bounded on $L^\infty(\Z^d)$.
  For each $t>0$, the integrand $\phi_t^*$ is continuous in $m^2$ and right-continuituous at $m^2=0$,
  and with the uniform bound \eqref{e:decomp-phi-est-j1},
  the claimed continuity of $C_1$
  follows from the continuity of $\phi_t^*$ by the dominated convergence theorem as for $j>1$.

  Next we verify the claims for $C_{N,N}$.
  Let $\varepsilon > 0$ and $m^2 \geq \varepsilon L^{-2(N-1)}$.
  For $j \ge N$,
  we have $1+m^2L^{2(j-1)} \ge m^2L^{2(j-1)} \ge \varepsilon L^{2(j-N)}$ and hence,
  with $\varepsilon$-dependent constant $c$,
  \begin{equation}
  (1+m^2L^{2(j-1)})^{-k-d} 
  \leq c (1+m^2L^{2(j-1)})^{-k} L^{-2d(j-N)}.
  \end{equation}
  By \eqref{e:decomp-phi-est} with $k$ replaced by $k+d$, and by \eqref{e:Cdef} and
  \eqref{e:CNNdef}, it therefore follows that
  \begin{align} \label{e:CNN-sum-bd}
    |\nabla_x^\alpha \nabla_y^\beta C_{N,N;x,y}|
    &\leq \sum_{j=N}^\infty \sum_{z\in \Z^d} |\nabla_x^\alpha \nabla_y^\beta C_{j;x,y+zL^N}|
    \nnb
    &\leq c (1+m^2L^{2(N-1)})^{-k} \sum_{j=N}^\infty L^{d(j-N)} L^{-2d(j-N)} L^{-(j-1)(2[\phi]+|\alpha|_1+|\beta|_1)}
    \nnb
    &\leq c (1+m^2L^{2(N-1)})^{-k} L^{-(N-1)(2[\phi]+(|\alpha|_1+|\beta|_1))}
    ,
  \end{align}
  where we have used the estimates
  \begin{equation}
    \sum_{z \in \Z^d} \1_{zL^N \leq O(L^j)} = O(L^{d(j-N)})
    \quad
    \text{and}
    \quad
    \sum_{j=N}^\infty L^{-(j-N)d}= \frac{1}{1-L^{-d}} \leq 2 \quad\text{(for $L \ge 2$)}.
  \end{equation}
  This shows that \eqref{e:scaling-estimate} holds also for $C_{N,N}$ if
  $m^2L^{2(N-1)} \geq \varepsilon$ and thus completes the proof of (a).

  To verify that $C_{N,N}$ is continuous in $m^2 \in (0,\bar m^2)$, let
  \begin{equation}
    C_{N,N;x,y}^M = \sum_{j=N}^M \sum_{z\in \Z^d} C_{j;x,y+zL^N}.
  \end{equation}
  This is a finite sum (due to the finite range of $C_j$)
  of $m^2$-continuous functions, and thus is
  continuous in $m^2 \in (0,\bar m^2)$.
  Analogously to \eqref{e:CNN-sum-bd}, it can be seen that,
  uniformly in $m^2 \in [\varepsilon L^{-2(N-1)},\bar m^2)$,
  \begin{align}
    |C_{N,N;x,y}
    - C_{N,N;x,y}^M|
    \to 0 \quad \text{as $M\to\infty$}.
  \end{align}
  As the uniform limit of a sequence of continuous functions,
  $C_{N,N;x,y}$ is thus continuous in $m^2 \in [\varepsilon L^{-2N},\bar m^2)$.
  Since $\varepsilon > 0$ is arbitrary,
  $C_{N,N;x,y}$ is therefore continuous in $m^2 \in (0,\bar m^2)$.
  This completes the proof of (b).

\smallskip\noindent
(c)
  We make several references to \cite{Baue13a}.
  By \cite[(1.37)--(1.38)]{Baue13a}, there exist
  $c>0$ and a function $\bar\phi \in C_c^\infty(\R^d)$ such that
  \begin{equation}
  \label{e:phi-scalinglimit}
    \phi_t^*(x,y;0)=(c/t)^{d-2}\bar\phi(c(x-y)/t) +O(t^{-(d- 1)})
  \end{equation}
  (due to a typographical error,
  $c^{d-2}$ is absent on the right-hand side of \cite[(1.37)]{Baue13a}).
  The function $\bar\phi$ is given in terms of another function $W_1$ in
  \cite[(3.17)]{Baue13a} as $\bar\phi(x) = \int_{\R^d}W_1(|\xi|^2)e^{ix\cdot \xi}d\xi$.
  By \cite[Lemma~2.2, (2.22)]{Baue13a}, $W_1(\lambda) = \varphi(\lambda^{1/2})$
  where $\varphi : [0,\infty) \to \R$ is a
  function such that $\int_0^\infty t \varphi(t)dt=1$ and
  such that its Fourier transform $\hat\varphi(k) = (2\pi)^{-1} \int_{\R} \varphi (t)
  e^{ikt}dt$ has support in $[-1,1]$ (we have chosen $C=1$ as in
  \cite[Remark~2.4]{Baue13a}).  Thus,
  \begin{align}
  \label{e:phi-scalinglimit-2}
    \phi_t^*(x,y;0)
    &=
    (c/t)^{d-2}\int_{\R^d} \varphi(|\xi|) e^{ic(x-y)\cdot \xi/t} d\xi
    +O(t^{-(d- 1)})
    \nnb
    & =
    (t/c)^2\int_{\R^d} \varphi(|\xi|t/c) e^{i(x-y)\cdot \xi} d\xi
    +O(t^{-(d- 1)}).
  \end{align}
  Set
  \begin{equation}
    \rho(s)
    =
    \left( \frac{s}{2c}\right)^2
    \varphi\left(\frac{s}{2c}\right) \frac{1}{s}.
  \end{equation}
  By definition,
  \begin{equation}
  \lbeq{cjhat}
    \hat c_j(\xi) = L^{2j}\hat c_0(L^j \xi)
     =
     \frac{1}{|\xi|^2}
     \int_{L^{j-1}|\xi|}^{L^j|\xi|}\rho(s)ds.
  \end{equation}
  For $j \ge 2$, as in \refeq{Cdef}, interchange of integration (and the
  change of variables $s=2t|\xi|$) gives
  \begin{align}
    \int_{\frac 12 L^{j-1}}^{\frac 12 L^j}
    \frac{dt}{t}
    (t/c)^2
    \int_{\R^d} d\xi \; \varphi(|\xi|t/c) e^{i(x-y)\cdot \xi}
    &=
    \int_{\R^d} \hat c_j(\xi)  e^{i(x-y)\cdot \xi} d\xi
    =
    c_j(x).
  \end{align}
  This completes the proof.
\end{proof}

\subsubsection{Bounds on coefficients}
\label{sec:Greeks}

We now prove two lemmas which provide
estimates for the coefficients of $\varphi_\pt$ (and hence $\Vpt$).
The coefficients were defined in Section~\ref{sec:frcd}, in terms of
the covariance decomposition
$(C_j)$ of $[-\Delta_\Zd + m^2]^{-1}$ given by \refeq{Cdef}.

\begin{lemma}
  \label{lem:wlims}
  Let $d \geq 4$, $j \ge 0$, $\bar m^2 >0$, $k \in \R$.
  The following bounds hold uniformly in $m^2 \in [0,\bar m^2]$
  (with constants which may depend on $L,\bar m^2$ but not on $j$):
  \begin{align} \lbeq{greekbds}
    \beta_j, \theta_j, \sigma_j , \zeta_j
    &= O(L^{-(d-4)j}(1+m^2L^{2j})^{-k}),
    \\
    \eta_j',\pi_j',\xi_j'
    &= O(L^{-(d-2)j}(1+m^2L^{2j})^{-k}),
    \\
    \delta_j[(w^2)^{(**)}]
    &= O(L^{-(d-6)j}(1+m^2L^{2j})^{-k}),
  \end{align}
  \begin{equation} \lbeq{w1bds}
    w_j^{(1)} = O(L^{2j}),
    \quad
    w_j^{(**)} = O(L^{4j}),
    \quad
    (w_j^2)^{(**)} = O(L^{2j})
    .
  \end{equation}
  Moreover, the left-hand sides of \eqref{e:greekbds}--\eqref{e:w1bds}
  are continuous in $m^2 \in [0,\bar m^2]$.
\end{lemma}

\begin{proof}
The continuity of the left-hand sides of \eqref{e:greekbds}--\eqref{e:w1bds} in $m^2$
is a consequence of their definitions together with the continuity of $C_j$ given
by Proposition~\ref{prop:Cdecomp}(b).
Thus it suffices to prove the estimates.

Fix $k\geq 0$.  Within the proof,
we set $M_j = (1+m^2L^{2j})^{-k}$,
and all constants may depend on $L$ but not on $j$.
We use the uniform bounds \eqref{e:scaling-estimate}
extensively without further comment.
With the finite-range property, they imply
\begin{equation} \label{e:Cconcl}
  |\nabla^l C_{j,x}|,
  |\nabla^l C_{j+1,x}|
  \leq O(M_j L^{-(d-2)j}L^{-lj}) \, \1_{|x|\leq O(L^{j})}, \quad l=0,1,2.
\end{equation}
The indicator functions in \eqref{e:Cconcl} give rise to volume factors in the estimates, i.e.,
\begin{equation}
  \sum_{x\in\Zd} \1_{|x|\leq O(L^{j})} \leq O(L^{dj}).
\end{equation}
We also frequently bound
a sum of exponentially growing terms by the largest term,
i.e., for $s > 0$,
\begin{equation}
  \sum_{l=1}^j L^{sl} \leq O(L^{sj}).
\end{equation}
Finally, we recall the definitions
\eqref{e:nuplusdef}--\eqref{e:wndef} with $w=w_j = \sum_{l=1}^j C_l$ and
 $C=C_{j+1}$.

\smallskip\noindent
\emph{Bound on $\beta_j$.}
By definition, $\beta_j$ is proportional to
\begin{equation}
  \delta[w^{(2)}] = 2(wC)^{(1)}+C^{(2)}.
\end{equation}
Using $dk-2[\phi]k=2k$ and $-2[\phi]j+2j = -(d-4)j$,
\begin{equation}
  (wC)^{(1)} = \sum_x C_{j+1,x} \sum_{k=1}^j C_{k,x} = O(M_j L^{-2[\phi]j}) \sum_{k=1}^j L^{dk}L^{-2[\phi]k} = O(L^{-(d-4)j}M_j),
\end{equation}
and similarly,
\begin{equation}
  C^{(2)} = \sum_x C_{j+1,x}^2 = O(M_j L^{dj} L^{-4j[\phi]}) = O(L^{-(d-4)j}M_j)
  .
\end{equation}

\smallskip\noindent
\emph{Bound on $\theta_j$.}
By definition, $\theta_j$ is proportional to
\begin{equation} \label{e:w3sssplit}
  \delta[(w^3)^{(**)}] = 3 (w^2C)^{(**)}+3(wC^2)^{(**)} + (C^3)^{(**)}
  .
\end{equation}
With $dk+2k-2[\phi]k=4k$ and $-4[\phi]j+4j = -2(d-4)j \leq -(d-4)j$,
\begin{align}
  (wC^2)^{(**)}
  =
  \sum_x \sum_{k=1}^j |x|^2 C_{k,x}C_{j+1,x}^2
  &\leq
  O(M_jL^{-4[\phi]j}) \sum_{k=1}^j L^{dk}L^{2k} L^{-2[\phi]k}
  \nnb
  &
  = O(L^{-(d-4)j}M_j),
\end{align}
and, with $-6[\phi]j+dj+2j = -(2d-4)j \leq -(d-4)j$,
\begin{equation}
  (C^3)^{**}
  =
  \sum_x |x|^2 C_{j+1,x}^3
  \leq O(M_j L^{-6[\phi]j} L^{dj}L^{2j}) \leq O(L^{-(d-4)j}M_j)
  .
\end{equation}
Also,
\begin{equation} \label{e:w2Csplit}
  (w^2C)^{(**)}
  =
  2\sum_x \sum_{l=1}^j\sum_{k=1}^{l-1} |x|^2 C_{k,x}C_{l,x}C_{j+1,x}
  +
  \sum_x \sum_{k=1}^j |x|^2 C_{k,x}^2 C_{j+1,x}.
\end{equation}
The first sum in \eqref{e:w2Csplit} is bounded, with $dk+2k-2[\phi]k=4k$ and $4l-2[\phi]l = (6-d)l$, by
\begin{align}
  \sum_x \sum_{l=1}^j\sum_{k=1}^{l-1} |x|^2 C_{k,x}C_{l,x}C_{j+1,x}
  &\leq
  O(M_jL^{-2[\phi]j})
  \sum_{l=1}^j\sum_{k=1}^{l-1} L^{dk}L^{2k}L^{-2[\phi]k} L^{-2[\phi]l}
  \nnb
  &\leq
  O(M_jL^{-2[\phi]j})
  \sum_{l=1}^j L^{(6-d)l}.
\lbeq{w2splitfirst}
\end{align}
The sum in \refeq{w2splitfirst} is bounded by $O(L^{2j})$ if $d\geq 4$ so that,
with $-2[\phi]j+2j = -(d-4)j$,
\begin{equation}
  \sum_x \sum_{l=1}^j\sum_{k=1}^{l-1} |x|^2 C_{k,x}C_{l,x}C_{j+1,x}
  \leq
  O(L^{-(d-4)j}M_j)
\end{equation}
as claimed.
The second term  in \eqref{e:w2Csplit} is similarly bounded,
with $dk+2k-4[\phi]k = (6-d)k$, as
\begin{equation}
  \sum_x \sum_{k=1}^j |x|^2 C_{k,x}^2 C_{j+1,x}
  \le
  O(M_jL^{-2[\phi]j})
  \sum_{k=1}^j L^{dk} L^{2k}L^{-4[\phi]k}
  \leq O(L^{-(d-4)j})M_j)
  .
\end{equation}
This completes the proof of \refeq{greekbds}.

\smallskip\noindent
\emph{Bound on $\eta_j'$.}
It follows immediately from \refeq{scaling-estimate} that
\begin{equation}
  \eta_j' = C_{j+1,0}
  \leq O(M_j L^{-(d-2)j}).
\end{equation}

\smallskip\noindent
\emph{Bound on $\xi_j'$.}
By definition, $\xi_j'$ is the sum of three terms.
The third term is
trivially bounded by $\eta_j'$.
The remaining two terms are proportional to
\begin{align} \label{e:delta-w3-sum}
    \delta[w^{(3)}] - 3w_j^{(2)}C_{j+1;0,0}
    &=
    \big(w_{j+1}^{(3)} - w_{j}^{(3)}\big)  - 3w_j^{(2)}C_{j+1;0,0}
    \nnb
    &=
    3 \left((w_{j}^{2}C_{j+1})^{(1)}  - w_j^{(2)}C_{j+1;0,0}\right) +
    3 (w_{j}C_{j+1}^{2})^{(1)} +
    C_{j+1}^{(3)}
    .
\end{align}
To bound the last two terms of \eqref{e:delta-w3-sum}, we use
$-6[\phi]j+dj = -2dj+6j \leq -(d-2)j$ to obtain
\begin{align}
  C_{j+1}^{(3)}
  =
  \sum_{y} C_{j+1,x}^3
  \le
  O(M_j L^{dj} L^{-6[\phi]j})
  \le
  O(L^{-(d-2)j}M_j).
\end{align}
Similarly, we use $dk-2[\phi]k=2k$ and
$-4[\phi]j+2j = -2dj+6j \leq -(d-2)j$ to obtain
\begin{align}
  (w_{j}C_{j+1}^{2})^{(1)}
  \le
  O(M_jL^{-4[\phi]j}) \sum_{k =1}^j  \sum_{x} C_{k,x}
  &\le
  O(M_jL^{-4[\phi]j})
  \sum_{k=1}^j L^{dk} L^{-2[\phi]k}
  \nnb
  &\le O(L^{-(d-2)j} M_j)
  .
\end{align}
The first term in \eqref{e:delta-w3-sum} is proportional to
\begin{align}
  \big(w_{j}^{2} (C_{j+1}-C_{j+1,0}) \big)^{(1)}
  =
  \sum_{k=0}^{j-1}
  \sum_{x}
  \delta_k[w_{x}^{2}]
  (C_{j+1,x} - C_{j+1,0}),
\end{align}
where we have used
\begin{equation}
  w_{j,x}^{2}
  =
  \sum_{k=0}^{j-1} \delta_k[w_{x}^{2}]
  \quad
  \text{with} \quad \delta_k[w_x^2] = w_{k+1,x}^2 - w_{k,x}^2.
\end{equation}
The bounds
\begin{align}
  \left|C_{j+1,x}-C_{j+1,0}-\sum_{i=1}^d x_i (\nabla_{i}C)_0\right|
  &\leq O(|x|^2 \|\nabla^2 C_{j+1}\|_\infty)
  \nnb
  &\leq O(M_j L^{-2[\phi]j}L^{-2j}) |x|^2
  ,
\end{align}
\begin{equation}
  \sum_x \delta_k[w_{x}^{2}]|x|^2
  = O(L^{2k}) \sum_x \delta_k[w_x^{2}] = O(L^{2k}\beta_k) = O(L^{2k})
  ,
\end{equation}
and the identity (which follows from $w_{-x}^2=w_x^2$)
\begin{equation}
  \sum_{x} \sum_{i=1}^d \delta_k[w_x^{2}] x_i (\nabla_i C)_0
  = - \sum_x \delta_k[w_x^{2}] x_i (\nabla_i C)_0 = 0
\end{equation}
then imply
\begin{equation}
  \big(w_{j}^{2} (C_{j+1}-C_{j+1,0}) \big)^{(1)}
  \le
  O(M_j L^{-2j[\phi]}L^{-2j})
  \sum_{k=0}^{j-1}
  L^{2k}
  =
  O(L^{-(d-2)j}M_j).
\end{equation}
This gives the desired bound on $\xi_j'$.

\smallskip\noindent
\emph{Bound on $\sigma_j$.}
By definition,
\begin{equation}
  \sigma
  =
  \delta[(w\Delta w)^{(**)}] = (C\Delta w)^{(**)} + (w\Delta C)^{(**)} + (C\Delta C)^{(**)}
  .
\end{equation}
Since $dk-2[\phi]k=2k$ and $-2[\phi]j+2j=-(d-4)j$,
\begin{equation}
  (C\Delta w)^{(**)}
  = O(L^{-2[\phi]j}M_j) \sum_{k=1}^j L^{dk} L^{2k} L^{-2k} L^{-2[\phi]k}
  = O(L^{-(d-4)j}M_j).
\end{equation}
Since $dk+2k-2[\phi]k=4k$ and $-2[\phi]j-2j+4j = -(d-4)j$,
\begin{equation}
  (w\Delta C)^{(**)}
  = O(L^{-2[\phi]j}L^{-2j}M_j) \sum_{k=1}^j L^{dk} L^{2k} L^{-2[\phi]k}
  = O(L^{-(d-4)j}M_j).
\end{equation}
Since $-4[\phi]j-2j+6j = -2dj + 8j \leq -(d-4)j$,
\begin{equation}
  (C\Delta C)^{(**)}
  = O(L^{-4[\phi]j}L^{-2j}L^{dj}L^{2j}M_j)
  = O(L^{-(d-4)j}M_j)
  .
\end{equation}
Together the above three estimates give the required result.

\smallskip\noindent
\emph{Bound on $\zeta_j$.}
The proof is analogous to the bound of $\sigma_j$ and is omitted.

\smallskip\noindent
\emph{Bound on $\pi_j'$.}
By definition, $\pi_j'$ is proportional to
\begin{equation}
  \delta[(w\Delta w)^{(1)}]
  =
  \sum_{k=1}^{j+1} \sum_{x}
  \big(C_{k,x} \,\Delta C_{j+1,x} +  \Delta C_{k,x} \,C_{j+1,x}\big)
  =
  2 \sum_{k=1}^{j+1} \sum_{x}
  C_{k} (x) \,\Delta C_{j+1,x}.
\end{equation}
With $dk-2[\phi]k = 2k$,
\begin{equation}
  \sum_{k=1}^{j+1} \sum_{x}
  C_{k} (x) \,\Delta C_{j+1,x}
  \le
  O( M_j L^{-2[\phi]j} L^{-2j}) \sum_{k =1}^{j+1}
  L^{dk} L^{-2[\phi]k}
  =O(L^{-(d-2)j}M_j)
  ,
\end{equation}
as required.

\smallskip\noindent
\emph{Bound on $\delta[(w^2)^{**}]$ and $(w^2)^{**}$.}
By definition,
\begin{equation}
  \delta[(w^2)^{**}] = 2(wC)^{(**)} + (C^2)^{(**)}.
\end{equation}
With $dk+2k-2[\phi]k=4k$ and $-2[\phi]j+4j = (6-d)j$,
\begin{equation}
  (wC)^{(**)}
  = \sum_{k=1}^j \sum_x |x|^2 C_{k,x}C_{j+1,x}
  = O(M_j L^{-2[\phi]j}) \sum_{k=1}^j L^{dk}L^{2k}L^{-2[\phi]k}
  = O(M_j L^{(6-d)j}),
\end{equation}
and similarly,
\begin{equation}
  (C^2)^{(**)}
  = \sum_x |x|^2 C_{j+1,x}^2
  = O(M_j L^{-4[\phi]j} L^{dj}L^{2j})
  = O(M_j L^{(6-d)j}).
\end{equation}
Since $6-d \leq 2$ for $d \geq 4$, taking the sum over $\delta_k[(w^2)^{(**)}]$
also implies the bound $(w^2)^{(**)} = O(L^{2j})$.

\smallskip\noindent
\emph{Bound on $w_j^{(1)}$.}
By definition,
\begin{equation}
  w^{(1)}
  = \sum_x \sum_{k=1}^j C_{k,x}
  = O(1)\sum_{k=1}^j L^{dk}L^{-2[\phi]k}
  = O(1)\sum_{k=1}^j L^{2k}
  = O(L^{2j}).
\end{equation}

\smallskip\noindent
\emph{Bound on $w_j^{(**)}$.}
By definition,
\begin{equation}
  w^{(**)}
  = \sum_x \sum_{k=1}^j |x|^2 C_{k,x}
  = O(1) \sum_{k=1}^j L^{dk}L^{2k}L^{-2[\phi]k}
  = O(1) \sum_{k=1}^j L^{4k}
    = O(L^{4j}).
\end{equation}

\smallskip\noindent
This completes the proof.
\end{proof}

\begin{lemma}
  \label{lem:betalim}
  (a) For $m^2 =0$, $\lim_{j \to \infty} \beta_j = 0$ for $d>4$, whereas
  \begin{equation}
    \label{e:delw2lim}
    \lim_{j \to \infty} \beta_j
    =
    \frac{\log L}{\pi^2}
    \quad\quad
    \text{for $d=4$}.
  \end{equation}
  (b)
  Let $d=4$ and $\bar m^2 >0$.
  There is a constant $c'$ independent of $j,L$ such that,
  for $m^2 \in (0,\bar m^2)$ and $j>1$,
  \begin{equation} \label{e:betaderiv}
    \left|\ddp{}{m^2} \beta_j(m^2)\right| \leq c'  (\log L) L^{d+2j}
    .
  \end{equation}
\end{lemma}

\begin{proof}
  (a)
  The conclusion for $d>4$ follows immediately from \refeq{greekbds},
  and we consider henceforth the case $d=4$.
  In this proof, constants
  in error estimates may depend on $L$.

  Let $c_0\in C_c(\R^4)$ be the function defined by
  Proposition~\ref{prop:Cdecomp}(c), and let $c_j(x) = L^{-{2j}}c_0(L^{-j}x)$,
  so that
  \begin{equation} \label{eq:app-C-approx}
    C_{j,x} = c_j(x) + O(L^{-3j}).
  \end{equation}
  We use the notation $(F,G) = \sum_{x\in\Z^4} F_xG_x$ for $F,G:\Z^4 \to \R$, and
  $\la f,g\ra = \int_{\R^4} fg \; dx$ for $f,g:\R^4\to\R$.
  We first verify that
  \begin{equation} \label{eq:Cc-approx}
    (C_{j}, C_{j+l}) - \la c_0, c_l \ra = O(L^{-j -2l}).
  \end{equation}
  Let $R_{j,x} = C_{j,x} - c_{j}(x)$. Then
  \begin{equation}
    (C_{j}, C_{j+l})
    = (c_{j},c_{j+l})
    + (c_{j}, R_{j+l}) + (c_{j+l}, R_j) + (R_j, R_{j+l}).
  \end{equation}
  Riemann sum approximation gives
  \begin{align}
    (c_{j},c_{j+l}) - \la c_{0},c_{l}\ra
    &= L^{-4j} \sum_{y\in L^{-j}\Z^d} c(y) c_l(y) - \int_{\R^d} c(y) c_l(y) \; dy\
    \nonumber\\
    &= O(L^{-j}) \|\nabla(c c_l)\|_{L^\infty}
    = 
    O(L^{-j-2l})
    .
  \end{align}
  The remaining terms are easily bounded using $|\supp(C_j)|,|\supp(R_j)| = O(L^{4j})$:
  \begin{alignat}{2}
    (c_{j},R_{j+l})
    &\leq O(L^{4j}) \|c_{j}\|_{L^\infty(\Z^4)} \|R_{j+l}\|_{L^\infty(\Z^4)}
    &&\leq O(L^{-j} L^{-3l}),
    \\
    (c_{j+l},R_{j})
    &\leq O(L^{4j}) \|c_{j+l}\|_{L^\infty(\Z^4)} \|R_{j}\|_{L^\infty(\Z^4)}
    &&\leq O(L^{-j} L^{-2l}),
    \\
    (R_j, R_{j+l})
    &\leq O(L^{4j}) \|R_j\|_{L^\infty(\Z^4)}\|R_{j+l}\|_{L^\infty(\Z^4)}
    &&\leq O(L^{-2j}L^{-3l}),
  \end{alignat}
  and \eqref{eq:Cc-approx} follows.

  From \eqref{eq:Cc-approx} we can now deduce that
  \begin{align} \label{eq:w2-v2-diff}
    \sum_{k=1}^{j} (C_k,C_{j+1})
    &= \sum_{k=1}^{j} \la c_0,c_{j+1-k} \ra + \sum_{k=1}^{j} O(L^{-k -2(j-k)})
    \nonumber\\ &
    = \sum_{k=1}^{j}  \la c_0,c_{k} \ra + O(L^{-j})
    ,
    \\
    (C_{j+1},C_{j+1})
    &= \la c_0,c_{0} \ra + O(L^{-j})
    .
  \end{align}
  Thus, using $\la c_0, c_k \ra = \la c_0, c_{-k} \ra$, we obtain
  \begin{equation}
    w_{j+1}^{(2)} -w_j^{(2)}
    = 2(w_j, C_{j+1}) + (C_{j+1}, C_{j+1})
    = \sum_{k = -j}^j \la c_0, c_k \ra  + O(L^{-j}).
  \end{equation}
  Application of $\|c_{-k}\|_{L^\infty} \leq L^{2k}\|c_0\|_{L^\infty}$
  and $\supp(c_{-k}) \subset B_{O(L^{-k})}$ gives
  \begin{align} \label{eq:betainftytail}
    \sum_{k=j+1}^{\infty} |\la c_0,c_{k}\ra|
    =
    \sum_{k=j+1}^{\infty} |\la c_0,c_{-k}\ra|
    &\leq
    \|c_0\|_{L^\infty} \sum_{k=j+1}^{\infty} L^{2k}  \int_{B_{O(L^{-k})}} |c_0(x)| \; dx
    \nonumber\\
    &\leq
    \|c_0\|_{L^\infty}^2 \sum_{k=j+1}^{\infty} O(L^{-2k})
    \leq
    O(L^{-2j})
    .
  \end{align}
  Thus we have obtained
  \begin{equation}
    \lbeq{betajinf}
    \beta_j = 8(w_{j+1}^{(2)} -w_j^{(2)}) =
    \beta_\infty
    + O(L^{-j})
    \quad
    \text{with}
    \quad
    \beta_\infty = 8\sum_{k=-\infty}^\infty \la c_0, c_k \ra.
  \end{equation}

  The constant $\beta_\infty$ is determined as follows.
  By \eqref{e:betajinf},
  \begin{equation}
    \beta_\infty = 8 \la c_0, v \ra
    \quad
    \text{with} \quad v = \sum_{k\in\Z} c_k.
  \end{equation}
  By Plancherel's theorem and \refeq{cjhat},
  \begin{align}
    \la c_0, c_k \ra
    &= \frac{1}{(2\pi)^4}
    \int_{\R^4}
    |\xi|^{-4}  \left( \int_{L^{-1}|\xi|}^{|\xi|} \rho(t) \, dt \right)
    \left( \int_{L^{k-1}|\xi|}^{L^k|\xi|} \rho(t) \, dt \right)\; d\xi,
  \end{align}
  and hence, by Fubini's theorem, radial symmetry, and $\int_0^\infty \rho \, dt = 1$,
  \begin{align}
    \la c_0, v\ra &=
    \frac{\omega_3}{(2\pi)^4}
    \int_0^\infty
    \left( \int_{L^{-1}r}^{r} \rho(t) \, dt \right) \; \frac{dr}{r}
    =
    \frac{\omega_3}{(2\pi)^4}
    \int_0^\infty \left( \int_{t}^{Lt} \; \frac{dr}{r} \right) \rho(t) \, dt,
  \end{align}
  where $\omega_3 = 2\pi^2$ is the surface measure of the $3$-sphere as
  a subset of $\R^4$.
  The inner integral in the last equation is equal to $\log L$.
  Thus, again using $\int_0^\infty \rho \, dt = 1$,  we find that
  \begin{equation}
    \beta_\infty
    = \frac{8\omega_3}{(2\pi)^4}
    \log L
    = \frac{\log L}{\pi^2}
  \end{equation}
  as claimed.

  \smallskip\noindent
  (b)
  In this proof, we set $d=4$, and constants are independent of $L$.
  We write $f' = \ddp{}{m^2} f$.  Using the notation of \refeq{wndef}, we have
  \begin{equation}
    \beta_j' = 16 ((wC)^{(1)})' + 8 (C^{(2)})'
    .
  \end{equation}
  By \eqref{e:scaling-estimate}--\eqref{e:decomp-deriv},
  \begin{equation}
    (C^{(2)})' = 2 \sum_x C_{j+1,x}'C_{j+1,x} \leq O(L^{-2j} \log L) O(L^{d(j+1)}) \leq O(L^{d} (\log L) L^{2j})
  \end{equation}
  and, similarly,
  \begin{equation}
    ((wC)^{(1)})'
    =
    \sum_x \sum_{k=1}^{j} (C_{k,x}C_{j+1,x}'+C_{k,x}'C_{j+1,x}).
  \end{equation}
  Again by \eqref{e:scaling-estimate}--\eqref{e:decomp-deriv},
  \begin{align}
    \sum_x \sum_{k=1}^j C_{k,x}C_{j+1,x}'
    &\leq O(\log L) \sum_{k=1}^j L^{dk} L^{-2(k-1)}
    \nnb &
    \leq O(L^d \log L) \sum_{k=1}^j L^{(d-2)(k-1)}
    \leq O(L^d (\log L) L^{2j}),
  \end{align}
  \begin{equation}
    \sum_x \sum_{k=1}^j C_{k,x}'C_{j+1,x} \leq O(L^{-2j} \log L) \sum_{k=1}^j O(L^{dk}) \leq O(L^{d} (\log L) L^{2j})
    .
  \end{equation}
  This completes the proof.
\end{proof}

\subsection{Proof of Proposition~\ref{prop:transformation}}
\label{sec:pf-transformation}

\begin{proof}[Proof of Proposition~\ref{prop:transformation}]
Let $\mu_+ = L^{2(j+1)}\nu_+$.
By \eqref{e:delta-def}, \refeq{newflow-g}--\refeq{newflow-mu} are equivalent to
(we drop superscripts $(0)$ on $z$ and $\zpt$)
\begin{align}
\lbeq{newflow-g-mu}
  [ g_{\pt} + 4 g \mu_+  \bar w_{j+1}^{(1)}]
  &=
  [ g   + 4 g  \mu  \bar w_j^{(1)}]
  -
  \beta_j  g^{2}
  ,
  \\[2mm]
  \lbeq{newflow-z-mu}
  [ z_{\pt}
  + 2  z  \mu  \bar w_{j+1}^{(1)}
  + \tfrac{1}{2}  \mu_+^{2}  \bar w_{j+1}^{(**)}]
  &=
  [ z
  + 2  z  \mu  \bar w_j^{(1)}
  + \tfrac{1}{2}  \mu^{2} \bar  w_j^{(**)}]
  - \theta_j  g^{2}
  ,
  \\[2mm]
  \lbeq{newflow-mu-mu}
  [ \mu_{\pt}  +  \mu_+^2  \bar w_{j+1}^{(1)}]
  &=
  L^2 [ \mu +   \mu^{2}  \bar w_j^{(1)}]
  + \eta_j [ g + 4 g  \mu \bar {w}_j^{(1)}]
  \nnb & \qquad
  -
  \xi_j  g^{2}
  -
  \omega_j
  g \mu
   - \pi   g z
  .
\end{align}
The form of the rewritten equations \eqref{e:newflow-g-mu}--\eqref{e:newflow-mu-mu}
suggests that we define maps $T_j:\R^3 \to \R^3$ by $T_j (g,z,\mu)= (\gch,\zch,\much)$ where
$(\gch,\zch,\much)$ are as in \eqref{e:gch-def1}--\eqref{e:much-def1}, i.e.,
\begin{align}
  \label{e:gch-def}
  \gch  & = g +4g\mu \bar w_j^{(1)},
  \\
  \label{e:zch-def}
  \zch & = z + 2z\mu \bar w_j^{(1)} + \tfrac{1}{2}  \mu^2 \bar w_j^{(**)},
  \\
  \label{e:much-def}
  \much & = \mu + \mu^2 \bar w_j^{(1)}
  .
\end{align}
By the inverse function theorem \cite[(10.2.5)]{Dieu69},
there exists a ball $B_\epsilon(0) \subset \R^3$
such that $T_j$ is an analytic diffeomorphism from $B_\epsilon(0)$ onto its image.
Note that $\epsilon$ can be chosen uniformly in $j$ and $m^2$  by
the uniformity of the bounds on $\bar w_j^{(1)}$ and
$\bar w_j^{(**)}$ in $j$ and $m^2$ of Lemma~\ref{lem:wlims}.
It also follows from the inverse function theorem that $T^{-1}(V) = V + O(|V|^2)$
with uniform constant.

The left-hand sides of \eqref{e:newflow-g-mu}--\eqref{e:newflow-mu-mu} equal
$T_{j+1}(\varphi_{\pt,j}^{(0)}(V)) + O((1+m^2L^{2j})^{-k}|V|^3)$ and the right-hand sides are
equal to $\bar\varphi_j(T_j(V)) + O((1+m^2L^{2j})^{-k}|V|^3)$.
For example,
with $T_{j+1}(V_\pt) = (\gch_\pt, \zch_\pt, \much_\pt)$,
it follows from Lemma~\ref{lem:wlims} that
\begin{align}
  g_{\pt} + 4 g \mu_+  \bar w_{j+1}^{(1)}
  &=
  [g_{\pt}
  + 4 g_\pt \mu_\pt  \bar w_{j+1}^{(1)}]
  + 4 (g-g_\pt) \mu_+ \bar w_{j+1}^{(1)}
  + 4 g_\pt(\mu_+-\mu_\pt) \bar w_{j+1}^{(1)}
  \nnb
  &=
  \gch_\pt
  + 4(\beta_j g^2 + 4g \delta[\mu \bar w^{(1)}]) \mu_+\bar w_{j+1}^{(1)}
  \nnb
  &\qquad
  + 4g_\pt (
  - 4 \eta_j  \bar w^{(1)} g\mu
  +
  \xi_j g^{2}
  +
  \omega_j
  g \mu  + \pi_j gz
  + \delta[\mu^{2} \bar w^{(1)}])
  \nnb
  &=
  \gch_\pt
  + O((1+m^2L^{2j})^{-k}|V|^3).
\end{align}
Thus
\begin{equation}
  T_{j+1}(\varphi^{(0)}_{\pt,j}(V)) = \bar\varphi_j(T_j(V)) + O((1+m^2L^{-2j})^{-k}|V|^3)
\end{equation}
as claimed.
\end{proof}

\subsection{Proof of Proposition~\ref{prop:rg-pt-flow}}
\label{sec:pf-alpha}

\begin{lemma} \label{lem:beta-jm}
  Let $d=4$, $\bar m^2 >0$, and $m^2 \in [0,\bar m^2]$.
  For any $c< \pi^{-2}\log L$, there exists $n < \infty$ such that
  $\beta_j(m^2) \geq c$ for $n \leq j \leq j_m-n$,
  uniformly in $m^2 \in [0,\bar m^2]$.
\end{lemma}

\begin{proof}
  Let $\varepsilon >0$ satisfy $c+\varepsilon < \pi^{-2}\log L$.
  By \eqref{e:delw2lim},
  there exists $n_0$ such that
  $\beta_j(0) \geq c+\varepsilon$ if $j \geq n_0$.
  This is sufficient for the case $m^2 =0$, where $j_m-n = \infty$.

  Thus we consider $m^2>0$.
  With $c'$ the constant in \refeq{betaderiv}, choose $n_1$ such that
  $c' (\log L)L^{4-2n_1} \le \varepsilon$.
  By definition in \eqref{e:jmdef}, $j_m = \lfloor \log_{L^2} m^{-2}\rfloor$,
  so $m^2 L^{2j_m} \le 1$.
  Thus, for $m^2\in (0,\bar m^2]$, \eqref{e:betaderiv} implies that
  if $1 <  j \leq j_m-n_1$ then
  \begin{equation}
    |\beta_j(0)-\beta_j(m^2)| \leq c'(\log L)L^{4+2j}m^2 \leq c'(\log L)L^{4-2n_1} \leq \varepsilon.
  \end{equation}
  Therefore, $\beta_j(m^2) \geq c$ if $n_0 \leq j \leq j_m-n_1$, as claimed.
\end{proof}

\begin{proof}[Proof of Proposition~\ref{prop:rg-pt-flow}]
  The continuity in $m^2 \in [0,\bar m^2]$
  of the coefficients in
  \refeq{gpt2a}--\refeq{qpt2} and
  in \eqref{e:newflow-gbar}--\eqref{e:newflow-mubar}
  is immediate from Proposition~\ref{prop:Cdecomp} and Lemma~\ref{lem:wlims}.

  To verify that $\bar\varphi$ obeys
  \cite[Assumptions~(A1--A2)]{BBS-rg-flow}, we fix $\Omega > 1$, and recall
  from \refeq{jOmegadef} the definition
  \begin{equation}
    j_\Omega = \inf \{ k \geq 0: |\beta_j| \leq \Omega^{-(j-k)} \|\beta\|_\infty
    \text{ for all $j$} \}
    .
  \end{equation}
  Let $k$ be such that $L^{2k} \geq \Omega$. Then, for all $j \ge 0$,
  \begin{equation} \label{e:mchibd}
    (1+m^2L^{2j})^{-k}
    \leq
    L^{-2k(j-j_m)_+}
    \leq \Omega^{-(j-j_m)_+}
    .
  \end{equation}
  Fix $c,n$ as in Lemma~\ref{lem:beta-jm}.  Since $j_m \to \infty$ as $m \downarrow 0$,
  there is a $\delta$ such that $j_m >n$ when $m^2 \in [0,\delta]$.  For such $m$, it follows
  from Lemma~\ref{lem:beta-jm} that $\|\beta\|_\infty \ge c$.
  We apply Lemma~\ref{lem:wlims} and \eqref{e:mchibd}
  to see that there is a constant $C$ such that
  \begin{equation}
    |\beta_j|
    \leq C \Omega^{-(j-j_m)_+}
    \leq \frac{C}{c} \Omega^{-(j-j_m)_+} \|\beta\|_\infty
    \leq \Omega^{-(j-k)_+} \|\beta\|_\infty
  \end{equation}
  whenever $k \geq j_m + \log_\Omega (C /c)$.
  In particular, $j_\Omega \leq k$
  and thus $j_\Omega \leq j_m+O(1)$.
  On the other hand, by Lemma~\ref{lem:beta-jm},
  $\beta_{j_m-n} \geq c$, and the definition of $j_\Omega$ thus requires that
  $c \leq \beta_{j_m-n} \leq \Omega^{-(j_m-n-j_\Omega)_+} \|\beta\|_\infty$.
  Therefore, $j_m-n-\jm \le \log_\Omega(c^{-1}\|\beta\|_\infty)$, which implies
  that $\jm \ge j_m-n - \log_\Omega( c^{-1}\|\beta\|_\infty)$.
  This completes the proof of \eqref{e:jmjOmega}.
  Also, the number of $j\leq j_\Omega$ with $\beta_j < c$ is bounded
  by $n+\log_\Omega (C /c)$.
  This proves \cite[Assumption~(A1)]{BBS-rg-flow} and also shows
  \begin{equation}
  \lbeq{mOmega}
    \Omega^{-(j-j_m)_+} = O(\Omega^{-(j-\jm)_+}).
  \end{equation}
  Then \cite[Assumption~(A2)]{BBS-rg-flow}
  follows from Lemma~\ref{lem:wlims}, \eqref{e:mchibd}, and the previous sentence.
  This completes the proof.
\end{proof}

Finally, for use in \cite{BS-rg-IE}, we note the following inequalities.
First, it follows from
Proposition~\ref{prop:rg-pt-flow} and
\cite[Lemma~\ref{flow-lem:elementary-recursion}]{BBS-rg-flow}
that the sequence $(\gbar_j)$ solving \refeq{newflow-gbar} obeys (for sufficiently small
$\gbar_0$)
\begin{equation}
  \label{e:gbarmono}
  \frac 12 \gbar_{j+1} \le
  \gbar_j \le 2 \gbar_{j+1}.
\end{equation}
In addition, the combination of \refeq{scaling-estimate}, \refeq{mchibd},
and \refeq{mOmega} implies that there is an $L$-independent constant $c$
such that for $m^2 \in [0, \delta]$ and $j=1,\ldots,N-1$, 
and in the special case $C_j=C_{N,N}$ for
$m^2 \in [\varepsilon L^{-2(N-1)},\delta]$
with the constant $c$ now depending on $\varepsilon>0$,
\begin{equation}
  \label{e:scaling-estimate-Omega}
  |\nabla_x^\alpha \nabla_y^\beta C_{j;x,y}|
  \leq c \Omega^{-(j-\jm)_+}
  L^{-(j-1)(2[\phi]+(|\alpha|_1+|\beta|_1))}.
\end{equation}

\section*{Acknowledgements}

This work was supported in part by NSERC of Canada.
RB gratefully acknowledges
the support and hospitality of the IAM at the University of Bonn, and
of the Department of Mathematics and Statistics at McGill University,
where part of this work was done.
DB gratefully acknowledges the support and hospitality of
the Institute for Advanced Study at Princeton and of Eurandom during part
of this work.
GS gratefully acknowledges the support and hospitality of
the Institut Henri Poincar\'e,
where part of this work was done.

\bibliography{../../bibdef/bib}
\bibliographystyle{plain}

\end{document}